\documentclass[referee]{gji}
\usepackage{flushend}
\usepackage{setspace}
\usepackage{timet}
\usepackage{color}
\usepackage{balance}
\usepackage{amsmath}    
\usepackage{amssymb}    
\usepackage{mathrsfs}  
\usepackage{graphicx}
\usepackage{subfigure}
\usepackage{epstopdf}
\usepackage{tikz}
\usetikzlibrary{graphs}
\usepackage{ulem}

\newcommand{\co}[1]{\color{blue} #1}
\newcommand{\coo}[1]{\color{red} #1}

\allowdisplaybreaks[4]
\title{Exact closed-form solutions for Lamb's problem (II): a moving point load}
\author[Xi Feng and Haiming Zhang]{Xi Feng and Haiming Zhang \\
  Department of Geophysics, School of Earth and Space Sciences, Peking University \emph{100871}, P. R. China.\\
  Email: zhanghm@pku.edu.cn}
\date{In original form 2020 March 28}
\pagerange{\pageref{firstpage}--\pageref{lastpage}}
\volume{200}
\pubyear{2020}

\graphicspath{{fig/}}
\begin{document}
\label{firstpage}

\maketitle

\begin{summary}
In this article, we report on an exact closed-form solution 
for the displacement in an elastic homogeneous half-space 
elicited by a downward vertical point source moving with constant velocity over the surface of the medium. 
The problem considered here is an extension to Lamb's problem.
Starting with the integral solutions of 
Bakker \textit{et al.}, 
we followed the method developed in Feng and Zhang, 
which focuses on the displacement triggered by a fixed point source observed on the free surface, 
to obtain the final solution 
in terms of elementary algebraic functions as well as elliptic integrals of the first, second and third kind.  
Our closed-form results agree perfectly with the numerical results of Bakker \textit{et al.}, 
which confirms the correctness of our formulas. 
The solution obtained in this article may lay a solid foundation for further consideration of the response of an actual physical moving load, such as a high-speed rail train.
\end{summary}

\section{Introduction}
The response of an isotropic homogeneous elastic half-space to a vertical load that moves with constant velocity on the surface of the half-space, which could be regarded as a natural extension of the classic Lamb's problem in seismology, has been the subject of numerous studies in the past half century.
Since the speed of high-speed railway train is close to that of seismic waves in the soil near the surface, the analysis on the controlling waves induced by these high-speed trains are essential nowadays.
The elastic half-space may be used as 
an ideal model with canonical geometry, because
it allows one to derive exact solutions to understand the property of the generated wave field. 
As such, the model provides the most fundamental physical insights into the transient response of a solid support to a moving load.

Early researches were focused on 2-D problems.
The solution of a line force moving steadily over the free surface was first presented by Sneddon ({\co 1952}), although the velocity of the source was limited to subseismic velocity.
Cole and Huth ({\co 1958}) extended Sneddon's result for all admissible velocities, and an error
in Cole and Huth's work was corrected by Georigiads and Barber (1993).
However, transient features of the moving load problem were not properly assessed in these papers, which were revealed by Ang ({\co 1960}) by means of the Fourier transform and the Cagniard-de Hoop method (Cagniard, {\co 1939}; de Hoop, {\co 1960}). 
Payton ({\co 1967}) also explored the surface motions elicited by the moving line load utilizing the dynamic Betti-Rayleigh reciprocal theorem 
(Sokolnikoff and Specht, {\co 1956}), and found that the displacement becomes unbounded when the source moves at the Rayleigh wave velocity. 
Freund ({\co 1972, 1973}) considered the non-uniform moving velocity for both steady and transient wave motion.

For the 3-D model, the steady-state motions by a moving point load were presented by Papadopoulos ({\co 1963}) and Eason ({\co 1965}), which focused on different ranges of the velocity of the load. Solutions for steady-state displacements at any point in the half-space and transient horizontal displacements on the surface were presented by Lansing ({\co 1965}). 
The closed-form expressions for the steady-state normal displacements were also derived by Barber ({\co 1996}). The Smirnov-Sobolev technique (Smirnov {\co 2014}) was used to reduce the problem to a linear superposition of two-dimensional stress and displacement fields. 
The static results for the tangential loads were developed in Georgiadis and Lykotrafitis ({\co 2001}), which were based on the use of the Radon transform and elements of distribution theory.
Moreover, a layered viscoelastic half-space was considered by de Barrors and Luco ({\co 1994}) to obtain the static displacements and stresses based on an integral representation in terms of the wavenumber. Recently, Kausel (2018) provide a numerical solution as a extension to the Stiffness Matrix Method (Kausel and Ro{\"e}sset, 1981), which is applicable for arbitrarily layered media, and arbitrary locations of the source and the receiver.

On the other hand, the transient response to a moving source has drawn much attention for richer wave field information involved.
The transient solutions for 3-D elastic half-space was first presented by Payton ({\co 1964}), although the displacements were confined to the free surface.
Gakenheimer and Miklowitz ({\co 1969}) extended Payton's ({\co 1964}) solutions and obtained transient motions in the interior of the medium for a normal point load, that is abruptly applied at a point and then traveled with a constant velocity along the free surface. 
The subseismic, transeismic and superseismic cases were all considered in their study with the inverse transform evaluated using the Cagniard-de Hoop technique, a very useful tool to obtain exact solutions directly in the time domain.
As Gakenheimer and Miklowitz's work could be used to solve more intricate problems, Bakker \textit{et al.} ({\co 1999}),
abbreviated as BVK99 hereafter, revisited their model using a simpler version of the Cagniard-de Hoop method. Bakker and Verweij ({\co 2012}) also focused on the knife-load (i.e. 2-D line load) with a similar procedure, extending the work of Norwood ({\co 1970}), who analyzed the semi-infinite line load moving at superseismic speed. These solutions were all obtained by various transform techniques, mainly the Fourier or Laplace transform, followed by analytical transform inversions.

Due to the complexity of the exact solution, some asymptotic solutions with simpler forms were introduced in recent years. De Hoop ({\co 2002}) assumed the horizontal displacements to be zero so as to determine closed-form expressions for the vertical displacement. Based on three kinds of simplified assumptions, Beskou \textit{et al.} ({\co 2018}) compared their solutions with exact solutions and analyzed their degrees of accuracy. 
A semi-analytical discretized mode was developed in Bierer and Bode ({\co 2007}) for the vertical component of the displacements at the fixed observation point at the surface. 
Another hyperbolic-elliptic model, taking advantage of the small parameter expressing the proximity of the load speed to the Rayleigh wave speed, were employed for 2-D medium (Kaplunov \textit{et al.}, {\co 2010}), 3-D medium (Kaplunov \textit{et al.}, {\co 2013}) and Gaussian-type profile distributed load (Ege and Erbaş, {\co 2017}), respectively.

When the velocity of the load reduces to zero, the problem reduces to the classic Lamb's problem, which traces back to Lamb's ({\co 1904}) pioneering work. A partial list of papers concerning the analytic solutions for 3-D problems can be found in Cagniard ({\co 1939}), and in Pekeris ({\co 1955a, 1955b}), Pekeris and Lifson ({\co 1957}), Chao ({\co 1960}), de Hoop ({\co 1960}) and Mooney ({\co 1974}). Johnson ({\co 1974}) collected these solutions together with a uniform notation. Starting with Johnson's formulae, an exact closed-form solution for Lamb's problem were eventually derived by Feng and Zhang ({\co 2018})
, which for the sake of brevity will be referred to as FZ18 hereinafter.
It is worth pointing out that the structure of the formulae in Johnson ({\co 1974}) and in BVK99 is quite similar. 
In both case the displacements are expressed in terms of definite integrals 
in the complex plane, which provided the starting point to our closed-form solutions for the moving point load problem.

This article presents the exact closed-form formulae for the displacement anywhere within a homogeneous, elastic half-space due to a vertical point load that moves with constant velocity over its surface. 
Starting with the integral expressions in BVK99, we succeeded in converting the integral forms into closed-form expressions, given in terms of standard elliptic integrals. 
Hence, our results have a clear advantage over their counterpart that must be solved by numerical integration with the waveforms are needed at miscellaneous receivers.
For instance, just the roots of the Rayleigh function are needed to capture the Rayleigh wave. 
Thus, exact algebraic operations substitute herin for the asymptotic methods usually used to calculate the numerical integrals. 
\section{Notation and definitions}
Consider a Cartesian coordinate system wherein the free surface is in the plane at $x_3=0$, while the positive $x_3$ axis points downward, as shown in Fig. 1.
$\lambda$ and $\mu$ are the Lam\'{e} constants of the elastic medium,
and $\rho$ is the density.
The source load is applied suddently at $t = 0$, at the origin
$O$, from where it starts moving along the $x_1$ axis with constant velocity $c$. 
We denote $r$ as the distance from the origin to the receiver, while $R$ is the projection of $r$ onto the horizontal $x_1 x_2$ plane. 
$\theta$ is the angle between $r$ and $R$, and $\phi$ between $R$ and $x_1$ axis.
\begin{figure}
\centering
\includegraphics[width=.55\textwidth]{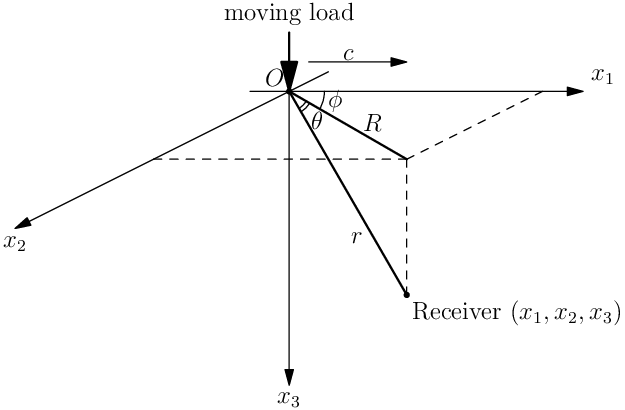}
\caption{The geometry of the problem. 
The Cartesian coordinate $x_1 x_2 x_3$ is used in this paper. 
The free surface of the homogeneous half-space is loaded with a vertical point force at $t=0$, which starts from the origin $O$ and propagates with a constant velocity $c$ along the $x_1$ axis. 
The coordinate of the receiver is ($x_{1}$, $x_{2}$, $x_3$). 
The distance between the origin and the receiver is $r$, and $R$ is the projection of $r$ on the place $x_1 x_2$.
$\theta$ is the angle between $r$ and $R$, and $\phi$ between $R$ and $x_1$ axis.}
\end{figure}
The following symbols are used throughout the paper:\\
\begin{tabular}{|c|l|}
\hline
$\lambda$, \,$\mu$ & Lam\'{e} constants \\
$\rho$ & mass density \\
$t$ & time \\
$\alpha$ & velocity of P wave \\
$\beta$ & velocity of S wave \\
$c$ & velocity of source \\
$k=\frac{\alpha}{\beta}$ & ratio of P to S wave velocity \\
$k_\alpha$ & $c/\alpha$\\
$k_\beta$ & $c/\beta$\\
$u$ & displacement\\
($x_{1}$, $x_{2}$, $x_{3}$) & coordinate of the receiver\\
\hline
\end{tabular}
\begin{tabular}{|c|l|}
\hline
$r$ & distance between source and receiver \\
$R$ & epicentral distance between source and receiver \\
$R_l$ & $\sqrt{x_2^2+x_3^2}$\\
$\theta$ & azimuth\\
$\phi$ & polar angle\\
$T_{\rmn P}=\tfrac{\alpha t}{r}$ & dimensionless time in $u^{\rm Pc}$ \\
$T_{\rmn S}=\tfrac{\beta t}{r}$ & dimensionless time in $u^{\rm Sc}$ and $u^{\rm S\text{-}Pc}$\\
${\rm sgn}(\cdot)$ & the sign function \\
$H(\cdot)$ &the Heaviside step function \\
${\rm Re}(\cdot)$ & real part \\
${\rm Im}(\cdot)$ & imaginary part \\
\hline
\end{tabular}
\section{Introduction to expressions in BVK99}
In this section, the integral expressions in BVK99, which is the starting point of our study, will be briefly reviewed.
In the original expressions in BVK99, the velocity of the moving source $c$ appears in the denominator, which will cause overflow in numerical experiments when $c$ approaches to zero. 
So their formulae should be rewritten to circumvent the problem.
The general displacement is separated into six parts as
\begin{align}\label{Full expression}
u_j=u_j^{\rm Pc}+u_j^{\rm Sc}+u_j^{\rm S\text{-}Pc}+
u_j^{\rm Pl}+u_j^{\rm Sl}+u_j^{\rm S\text{-}Pl}. \quad \ (j=1,2,3)
\end{align} 
The first three terms $u_j^{\rm Pc}$, $u_j^{\rm Sc}$ and $u_j^{\rm S\text{-}Pc}$ are associated respectively with the P wave, the S wave and the diffracted S-P wave in the half-space, and contain the contour integrations in the complex plane. The superscript $c$ denotes the contour of the integral. 
The last three terms $u_j^{\rm Pl}$, $u_j^{\rm Sl}$ and $u_j^{\rm S\text{-}Pl}$ are due to the residues at the pole within integral path, as indicated by the superscript $l$.  
The explicit expressions of $u_j^{\rm Pc}$, $u_j^{\rm Sc}$ and $u_j^{\rm S\text{-}Pc}$ together with their corresponding variables are listed in the ensuing:
\begin{align}
\label{Bakker uPc}
u_j^{\rm Pc}=\frac{1}{\pi^{2}\mu r}
&
\left.
\int^{p_{_{\rm P}}(t)}_{0}
H\left(T_{\rm P}-1\right)
{\rm Re}\left[ 
\frac{\eta_\alpha\mathcal{E}_{j}^{\rm P}(p,q)}
{W\sigma\sqrt{p^2_{_{\rm P}}(t)-p^2}}
\right]
\right|_{q=q_{_{\rm P}}(t,p)}\mathrm{d}p, \\
\label{Bakker uSc}
\notag 
u_j^{\rm Sc}=\frac{1}{\pi^{2}\mu r}
&
\left\{
\left.
\int^{p_{_{\rm S}}(t)}_{0}
H\left(T_{\rm S}-1\right)
{\rm Re}\left[ 
\frac{\eta_{_\beta}\mathcal{E}_{j}^{\rm S}(p,q)}
{W\sigma\sqrt{p^2_{_{\rm S}}(t)-p^2}}
\right]
\right|_{q=q_{_{\rm S}}(t,p)}\mathrm{d}p 
\right. 
\\
&\left.
+H\left(k\sin\theta-1\right)
\left.
\int_{\beta^{-1}\sqrt{T_{\rm S}^2-1}}^{p_{_{\rm S\text{-}P}}(t)}
H\left(T_{\rm S}-1\right)
{\rm Im}\left[
\frac{\eta_{_\beta}\mathcal{E}_{j}^{\rm S}(p,q)}
{W\sigma\sqrt{p^2-p^2_{_{\rm S}}(t)}}
\right]
\right|_{q=q_{_{\rm S\text{-}P}}(t,p)}\mathrm{d}p\right\}, 
\end{align}
\begin{small}
\begin{align}
\label{Bakker uSPc}
u_j^{\rm S\text{-}Pc}=-\frac{1}{\pi^{2}\mu r}
&\left.
H\left(k\sin\theta-1\right)
\int^{p_{_{\rm S\text{-}P}}(t)}_{0}
\left[
H\left(T_{\rm S\text{-}P}-1\right)
-H\left(T_{S}-1\right)\right]
{\rm Im}\left[
\frac{\eta_{_\beta}\mathcal{E}_{j}^{\rm S}(p,q)}
{W\sigma\sqrt{p^2-p^2_{_{\rm S}}(t)}}
\right]
\right|_{q=q_{_{\rm S\text{-}P}}(t,p)}\mathrm{d}p,
\end{align}
\end{small}
where
\begin{align*}
&p_{_{\rm P}}(t)=\alpha^{-1}\sqrt{T_{\rm P}^2-1},
&&
p_{_{\rm S}}(t)=\beta^{-1}\sqrt{T_{\rm S}^2-1},
\\
&
T_{\rm S\text{-}P}=T_{\rm P}\left(\sin\theta+\sqrt{k^2-1}\cos\theta\right)^{-1},
&& 
p_{_{\rm S\text{-}P}}(t)=\alpha^{-1}
\sqrt{\left(T_{\rm P}-\sqrt{k^2-1}\cos\theta\right)^2\sin^{-2}\theta-1}, 
\\
&
\eta_\alpha\left(p,q\right)=\sqrt{\alpha^{-2}+p^2-q^2},\quad 
({\rm Re}\left\{\eta_\alpha\right\}>0)  
&&
\eta_{_\beta}\left(p,q\right)=\sqrt{\beta^{-2}+p^2-q^2},\quad
({\rm Re}\left\{\eta_\beta\right\}>0)  
\\
&
W\left(p,q\right)=\left(cq-\tfrac{R}{x_1}\right)^2+c^2 \tfrac{x_2^2}{x_1^2}p^2,
&&
\sigma\left(p,q\right)=\left(\eta_{_\beta}^2+p^2-q^2\right)^2+4\eta_\alpha\eta_{_\beta}\left(q^2-p^2\right),
\end{align*}
and
\begin{align*}
q_{_{\rm P}}(t,p)&=\frac{t}{r}\sin{\theta}+{\rm i}\sqrt{p^2_{_{\rm P}}(t)-p^2}\cos\theta, \notag \\
q_{_{\rm S}}(t,p)&=\frac{t}{r}\sin{\theta}+{\rm i}\sqrt{p^2_{_{\rm S}}(t)-p^2}\cos\theta, \notag \\
q_{_{\rm S\text{-}P}}(t,p)&=\frac{t}{r}\sin{\theta}-\sqrt{p^2-p^2_{_{\rm S}}(t)}\cos\theta,
\end{align*}
\begin{align*}
&
\mathcal{E}_{1}^{\rm P}\left(p,q\right)=
-\gamma 
\left(cq^{2}+cp^2\tan^2\phi
-q\cos^{-1}\phi\right),
&&
\mathcal{E}_{1}^{\rm S}\left(p,q\right)=
2\eta_{\alpha}\eta_{\beta}
\left(cq^{2}
+cp^2\tan^2\phi
-q\cos^{-1}\phi\right),
\\
&
\mathcal{E}_{2}^{\rm P}\left(p,q\right)=
-\gamma 
\left(cq^{2}-cp^{2}
-q\cos^{-1}\phi\right)\tan\phi,
&&
\mathcal{E}_{2}^{\rm S}\left(p,q\right)=
2\eta_{\alpha}\eta_{\beta}
\left(cq^{2}-cp^{2}-q\cos^{-1}\phi\right)\tan\phi,
\\
&
\mathcal{E}_{3}^{\rm P}\left(p,q\right)=
-\gamma\eta_{\alpha}
\left(cq-\cos^{-1}\phi\right)\cos^{-1}\phi,
&&
\mathcal{E}_{3}^{\rm S}\left(p,q\right)=
2\eta_{\alpha}
\left(p^2-q^2\right)
\left(cq-\cos^{-1}\phi\right)\cos^{-1}\phi,
\end{align*}
where $\gamma\left(p,q\right)=\eta_{\beta}^2+p^2-q^2$.
The explicit expressions of $u_j^{\rm Pl}$, $u_j^{\rm Sl}$ and $u_j^{\rm S\text{-}Pl}$ with the corresponding variables are listed as follows:
\begin{align}
\label{Bakker uPl}
u_j^{\rm Pl}&=\frac{1}{\pi\mu}
H\left(T_{\rm Pl}-1\right)
{\rm Re}
\left.
\left[
\frac{\eta_{\alpha}\mathcal{E}^{\rm P}}
{\sigma\sqrt{\left(ct-x_1\right)^2-R_l^2a_{pc}^2}}
\right]
\right|_{q=q_{_{\rm P}}(t)},
\\
\label{Bakker uSl}
u_j^{\rm Sl}&=\frac{1}{\pi\mu}
H\left(T_{\rm Sl}-1\right)
{\rm Re}
\left.
\left[
\frac{\eta_{\beta}\mathcal{E}^{\rm S}}
{\sigma\sqrt{\left(ct-x_1\right)^2-R_l^2a_{sc}^2}}
\right]
\right|_{q=q_{_{\rm S}}(t)},
\\
\label{Bakker uSPl}
u_j^{\rm S\text{-}Pl}&=\frac{1}{\pi\mu}
H\left(T_{\rm S\text{-}Pl}-1\right)
{\rm Re}
\left.
\left[
\frac{\eta_{\beta}\mathcal{E}^{\rm S}}
{\sigma\sqrt{\left(ct-x_1\right)^2-R_l^2a_{sc}^2}}
\right]
\right|_{q=q_{_{\rm S\text{-}P}}(t)},
\end{align}
where
\begin{align*}
T_{\mathrm{Pl}}
&=\left\{\begin{array}{ll}
{\frac{ctx_1}{r^2},} 
& {\tfrac{r^2}{x_1^2}-k_{\alpha}^2>0} \\
{\frac{ct}{x_1}-\tfrac{R_l}{x_1}a_{pc},} 
& {\tfrac{r^2}{x_1^2}-k_{\alpha}^2 \leq 0}
\end{array}\right.
\\
T_{\mathrm{Sl}}
&=\left\{\begin{array}{ll}
{\frac{ctx_1}{r^2},} 
& {\tfrac{r^2}{x_1^2}-k_{\beta}^2>0} \\
{\frac{ct}{x_1}-\tfrac{R_l}{x_1}a_{sc},} 
& {\tfrac{r^2}{x_1^2}-k_{\beta}^2 \leq 0}
\end{array}\right.\\
T_{\mathrm{S\text{-}Pl}}
&=\left\{\begin{array}{ll}
{\frac{ctx_1}{R^2}-\frac{x_3}{R}\sqrt{k_{\beta}^2\cos^2\phi-1},} 
& {\tfrac{R^2}{x_1^2}-k_{\beta}^2>0} \\
{\frac{ct}{x_1}-a_{pc}\tan\phi-\tfrac{x_3}{x_1}\sqrt{k^2_\beta-k^2_\alpha},} 
& {\tfrac{R^2}{x_1^2}-k_{\beta}^2 \leq 0}
\end{array}\right.
\end{align*}
\begin{align*}
& 
a_{pc}=\sqrt{k^2_{\alpha}-1}, 
&& 
a_{sc}=\sqrt{k^2_{\beta}-1},
\\&
\eta_{\alpha}\left(q\right)=\sqrt{-\left(q+\tan\phi\right)^2+a_{pc}^2}, 
&&
\eta_{\beta}\left(q\right)=\sqrt{-\left(q+\tan\phi\right)^2+a_{sc}^2}, 
\\&
\gamma\left(q\right)=k_\beta^2-2-2\left(q+\tan\phi\right)^2,
&&
\sigma\left(q\right)=\gamma^2
+4\eta_\alpha\eta_{\beta}
\left[1+\left(q+\tan\phi\right)^2\right],
\end{align*}
and
\begin{align*}
q_{_{\rm P}}(t)&=\frac{1}{R_l^2}
\left[
x_2\left(ct-\tfrac{r^2}{x_1}\right)
+{\rm i}x_3\sqrt{\left(ct-x_1\right)^2-R_l^2a_{pc}^2}
\right], 
\\
q_{_{\rm S}}(t)&=\frac{1}{R_l^2}
\left[
x_2\left(ct-\tfrac{r^2}{x_1}\right)
+{\rm i}x_3\sqrt{\left(ct-x_1\right)^2-R_l^2a_{sc}^2}
\right], 
\\
q_{_{\rm S\text{-}P}}(t)&=\frac{1}{R_l^2}
\left[
x_2\left(ct-\tfrac{r^2}{x_1}\right)
-x_3\sqrt{R_l^2a_{sc}^2-\left(ct-x_1\right)^2}
\right], 
\end{align*}
\begin{align*}
&
\mathcal{E}_{1}^{\rm P}\left(q\right)=\gamma,
&&
\mathcal{E}_{1}^{\rm S}\left(q\right)=-2\eta_{\alpha}\eta_{\beta},
\\
&
\mathcal{E}_{2}^{\rm P}\left(q\right)=\gamma\left(q+\tan\phi\right),
&&
\mathcal{E}_{2}^{\rm S}\left(q\right)=-2\eta_{\alpha}\eta_{\beta}
\left(q+\tan\phi\right),
\\
&
\mathcal{E}_{3}^{\rm P}\left(q\right)=\gamma\eta_{\alpha},
&&
\mathcal{E}_{3}^{\rm S}\left(q\right)=
2\eta_{\alpha} 
\left[1+\left(q+\tan\phi\right)^2\right].
\end{align*}
In the next three sections,
$u_j^{\rm Pc}$, $u_j^{\rm Sc}$ and $u_j^{\rm S\textit{-}Pc}$ are investigated specifically
to express the results in terms of elementary functions and elliptic functions instead of the integral representations.
\section{Calculation of {$u_{\lowercase{j}}^{\rm P\lowercase{c}}$}}
In this section, $u_j^{\rm Pc}$ will be decomposed into elementary functions and elliptic integrals step by step. The procedures are quite similar to those in FZ18. And the detailed guide to implement our results are displayed in the summary in this section.

In order to transform equation ({\ref{Bakker uPc}}) into a form easy to integrate, a new variable $B=\alpha\eta_{\alpha}$ is introduced so that
\begin{align*}
&p=\displaystyle\frac{\sqrt{B^{2}-2T_{\rm P}B\cos{\theta}+T_{\rm P}^{2}-\sin^{2}{\theta}}}{\alpha\sin{\theta}}, && \displaystyle q=\frac{B\cos{\theta}-T_{\rm P}}{\alpha\sin{\theta}}.
\end{align*}
By these substitutions, equation ({\ref{Bakker uPc}}) can be transformed into
a standard integral form as
\begin{align}
\label{standard uPc}
u_j^{\rm Pc}&=\frac{1}{\pi^2\mu r}H\left(T_{\rm P}-1\right)
\int_{T_{\rm P}\cos\theta}^{T^{\rm P}_{\rm up}}
{\rm Im}\left(
\frac{M_j(B)}
{\sigma^{\rm P}\left(B\right)W^{\rm P}\left(B\right)}
\frac{{\rm d}B}{\sqrt{Q^{\rm P}_1\left(B\right)}}
+\frac{N_j(B)}
{\sigma^{\rm P}\left(B\right)W^{\rm P}\left(B\right)}
\frac{{\rm d}B}{\sqrt{Q^{\rm P}_2\left(B\right)}}
\right),
\end{align}
where
\begin{align}\label{uPc parameter1}
\notag
M_j(B)&=B\left(2B^2+k^2-2\right)^3 P_j\left(B\right),\\
\notag
N_j(B)&=4B^2\left(B^2-1\right)\left(B^2+k^2-1\right)\left(2B^2+k^2-2\right)P_j\left(B\right),
\\
P_1(B)&=\tfrac{x_1^2}{r^2}\left(k_{\alpha} B^2-\tfrac{x_3}{x_1} B+T_{\rm P}\tfrac{r}{x_1}-k_{\alpha}\right)
-k_{\alpha}\left(B^2-2T_{\rm P}B\cos\theta+T_{\rm P}^2-\sin^2\theta\right),
\\
\notag
P_2(B)&=\tfrac{x_1 x_2}{r^2}\left(k_{\alpha} B^2-\tfrac{x_3}{x_1} B+T_{\rm P}\tfrac{r}{x_1}-k_{\alpha}\right),
\\
\notag
P_3(B)&=\tfrac{x_1}{r}\left(k_{\alpha}B^2\cos\theta+\tfrac{R}{x_1}B\sin\theta
-k_{\alpha} T_{\rm P}B\right),
\end{align}
and
\begin{align}\label{uPc parameter2}
\notag
&T^{\rm P}_{\rm up}
=T_{\rm P}\cos{\theta}+{\rm i}\sqrt{T_{\rm P}^{2}-1}\sin{\theta}, 
\\
&\sigma^{\rm P}(B)
=(2B^{2}+k^{2}-2)^{4}-16B^{2}(B^{2}+k^{2}-1)(B^{2}-1)^{2},
\\
\notag
&W^{\rm P}(B)
=\left(k_{\alpha}B\cos\phi \cos\theta+\sin\theta
-k_{\alpha}T_{\rm P}\cos\phi\right)^2+k^2_{\alpha}\sin^2\phi
\left(B^2-2T_{\rm P}B\cos\theta+T^2_{\rm P}-\sin^2\theta\right),
\\
\notag
&Q^{\rm P}_1(B)
=B^{2}-2T_{\rm P}B\cos{\theta}+T_{\rm P}^{2}-\sin^{2}{\theta}, 
\\
\notag
&Q^{\rm P}_2(B)
=(B^{2}+k^{2}-1)(B^{2}-2T_{\rm P}B\cos{\theta}+T_{\rm P}^{2}-\sin^{2}{\theta}),
\end{align}
The goal in this section is to convert the standard integral 
(eq. (\ref{standard uPc})) into a closed-form expression, which is an algebraic expression relates to a mathematical expression or equation in which a finite number of symbols is combined using only the operations of addition, subtraction, multiplication, division, and exponentiation with constant rational exponents.
Our method is based on such two theorems (Armitage and Eberlein, {\co 2006}):
\begin{enumerate}
\item An integral of the type $\displaystyle \int R(x, y){\mathrm d}x$, where $y^2$ is a quadratic polynomial in $x$, and $R$ denotes a rational function of $x$ and $y$, is called an elementary integral, which can be expressed as a finite sum of elementary functions.
\item An integral of the type $\displaystyle \int R(x, y){\mathrm d}x$, where $y^2$ is a quartic polynomial in $x$, and $R$ denotes a rational function of $x$ and $y$, is called an elliptic integral, which can be expressed as a finite sum of elementary functions and the three types of standard elliptic integral. 
\end{enumerate}
The three types of standard elliptic integrals are defined as
\begin{align*}
K\left(\tau\right)
&=\int_{0}^{1}\frac{\mathrm{d}x}{\sqrt{1-x^{2}}\sqrt{1-\tau^{2}x^{2}}},
\\
E\left(\tau\right)
&=\int_{0}^{1}\frac{\sqrt{1-\tau^{2}x^{2}}}{\sqrt{1-x^{2}}}\mathrm{d}x,
\\
\varPi\left(\tau,c\right)
&=\int_{0}^{1}\frac{1}{1-cx^{2}}\frac{\mathrm{d}x}{\sqrt{1-x^{2}}\sqrt{1-\tau^{2}x^{2}}},
\end{align*}
where $0<\tau<1$ and $c<1$. 
$K\left(\tau\right)$, $E\left(\tau\right)$ and $\varPi\left(\tau,c\right)$ are called elliptic integral of the first kind, second kind and third kind, respectively, which are calculated by commands ``ellipticK'', ``ellipticE'' and ``ellipticPi'' in Matlab, respectively.  

Following the theorems, the integration containing $\sqrt{Q^{\rm P}_1(B)}$ is an elementary integral and the integration containing $\sqrt{Q^{\rm P}_2(B)}$ is an elliptic integral. 
The next standard step is to decompose the fractions $\frac{M_j\left(B\right)}{\sigma^{\rm P}(B)W^{\rm P}(B)}$ and $\frac{N_j\left(B\right)}{\sigma^{\rm P}(B)W^{\rm P}(B)}$ into a series of monomials and partial fractions as 
\begin{align}
\notag
\frac{M_j\left(B\right)}{\sigma^{\rm P}(B)W^{\rm P}(B)}
=&\sum_{s=1}^{4}\frac{u^{\rm P}_{j,s}}{B-\sigma^{\rm P}_{s}}
+\frac{u^{\rm P}_{j,5}}{B^2+\left({\rm Im}\left(\sigma^{\rm P}_{5}\right)\right)^2}
+\frac{u^{\rm P}_{j,6}B}{B^2+\left({\rm Im}\left(\sigma^{\rm P}_{6}\right)\right)^2}
+\frac{u^{\rm P}_{j,7}}{\left(B-{\rm Re}\left(\omega^{\rm P}_1\right)\right)^2
+\left({\rm Im}\left(\omega^{\rm P}_{1}\right)\right)^2}
\\&
\label{elementary fraction}
+\frac{u^{\rm P}_{j,8}B}{\left(B-{\rm Re}\left(\omega^{\rm P}_2\right)\right)^2
+\left({\rm Im}\left(\omega^{\rm P}_{2}\right)\right)^2}
+u^{\rm P}_{j,9}+u^{\rm P}_{j,10}B,
\\
\notag
\frac{N_j\left(B\right)}{\sigma^{\rm P}(B)W^{\rm P}(B)}
=&\sum_{s=1}^{4}\frac{v^{\rm P}_{j,s}}{B-\sigma^{\rm P}_{s}}
+\frac{v^{\rm P}_{j,5}}{B^2+\left({\rm Im}\left(\sigma^{\rm P}_{5}\right)\right)^2}
+\frac{v^{\rm P}_{j,6}B}{B^2+\left({\rm Im}\left(\sigma^{\rm P}_{6}\right)\right)^2}
+\frac{v^{\rm P}_{j,7}}{\left(B-{\rm Re}\left(\omega^{\rm P}_1\right)\right)^2
+\left({\rm Im}\left(\omega^{\rm P}_{1}\right)\right)^2}
\\&
\label{elliptic fraction}
+\frac{v^{\rm P}_{j,8}B}{\left(B-{\rm Re}\left(\omega^{\rm P}_2\right)\right)^2
+\left({\rm Im}\left(\omega^{\rm P}_{2}\right)\right)^2}
+v^{\rm P}_{j,9}+v^{\rm P}_{j,10}B+v^{\rm P}_{j,11}B^2.
\end{align}
The coefficients in equation $u^{\rm P}_{j,i}\ (i=1,\cdots,10)$ and $v^{\rm P}_{j,i}\ (i=1,\cdots,11)$ are called the elementary coefficients and the elliptic coefficients for the same reason, respectively. 
$\sigma^{\rm P}_i\ (i=1,2,\cdots,6)$ are roots of $\sigma^{\rm P}(B)$.
$\sigma^{\rm P}_5$ and $\sigma^{\rm P}_6$ are always pure imaginary numbers, which are associated with the generation of the Rayleigh wave (Liu \textit{et al.}, {\co 2016}; FZ18), hence $\sigma^{\rm P}(B)$ is called the \textit{Rayleigh function}. More features of the Rayleigh wave are analyzed in Section 7.

The Poission ratio $\nu$ controls the distribution of other roots 
$\sigma^{\rm P}_i\ (i=1,2,\cdots,4)$, which are all real for $0<\nu<0.2631$,
and they are turned to complex for other values of $\nu$. For latter the corresponding expressions are somewhat cumbersome. 
Although the range of the Poission ratios is confined to $[0, 0.2631]$ in this paper, it is worth to emphasize that the main procedures are still valid for other values of $\nu$.
Fortunately, in the theoretical seismology researchers usually focus on the case with a Poisson's ratio 0.25, which is an adequate approximate value for the Earth (Lapwood, {\color{blue}1949}).

$W^{\rm P}(B)$ always has two conjugate roots, named $\omega^{\rm P}_{1}$ and 
$\omega^{\rm P}_{2}$. The behavior of $W^{\rm P}(B)$ depend on the velocity of the source $c$, and the polynomial will degenerate to a constant when $c$ decreases to zero. At this situation, results in FZ18 must be used to calculate the displacements as $\omega^{\rm P}_{1}$ and $\omega^{\rm P}_{2}$ vanish. 
All the coefficients could be easily evaluated by applying the Matlab command ``residue''. 

Apparently the whole integrations have been separated as the linear combinations of several simpler integrals $U_i^{\rm P}$ and $V_i^{\rm P}\ (i=1,2,\cdots,6)$, which containing $\sqrt{Q^{\rm P}_1(B)}$ and $\sqrt{Q^{\rm P}_2(B)}$, respectively,
\begin{align}\label{final uPc}
\notag
u^{\rm Pc}_{j}=&\frac{1}{\pi^{2}\mu r}
H\left(T_{\rm P}-1\right)
\Big[
\sum_{s=1}^4 u^{\rm P}_{j,s}U^{\rm P}_{1}\left(\sigma^{\rm P}_s\right)
+u^{\rm P}_{j,5}U^{\rm P}_{2}\left({\rm Im}\left(\sigma^{\rm P}_{5}\right), 0\right)
+u^{\rm P}_{j,6}U^{\rm P}_{3}\left({\rm Im}\left(\sigma^{\rm P}_{6}\right), 0\right)
\\&
\notag
+u^{\rm P}_{j,7}U^{\rm P}_{2}\left({\rm Im}\left(\omega^{\rm P}_{1}\right), -{\rm Re}\left(\omega^{\rm P}_{1}\right)\right)
+u^{\rm P}_{j,8}U^{\rm P}_{3}\left({\rm Im}\left(\omega^{\rm P}_{2}\right), -{\rm Re}\left(\omega^{\rm P}_{2}\right)\right)
+u^{\rm P}_{j,9}U^{\rm P}_{4}
+u^{\rm P}_{j,10}U^{\rm P}_{5}
\Big]
\\&
\notag
+\frac{1}{\pi^{2}\mu r}
H\left(T_{\rm P}-1\right)
\Big[
\sum_{s=1}^4 v^{\rm P}_{j,s}V^{\rm P}_{1}\left(\sigma^{\rm P}_s\right)
+v^{\rm P}_{j,5}V^{\rm P}_{2}\left({\rm Im}\left(\sigma^{\rm P}_{5}\right), 0\right)
+v^{\rm P}_{j,6}V^{\rm P}_{3}\left({\rm Im}\left(\sigma^{\rm P}_{6}\right), 0\right)
\\&
+v^{\rm P}_{j,7}V^{\rm P}_{2}\left({\rm Im}\left(\omega^{\rm P}_{1}\right), -{\rm Re}\left(\omega^{\rm P}_{1}\right)\right)
+v^{\rm P}_{j,8}V^{\rm P}_{3}\left({\rm Im}\left(\omega^{\rm P}_{2}\right), -{\rm Re}\left(\omega^{\rm P}_{2}\right)\right)
+v^{\rm P}_{j,9}V^{\rm P}_{4}
+v^{\rm P}_{j,10}V^{\rm P}_{5}
+v^{\rm P}_{j,11}V^{\rm P}_{6}\Big],
\end{align}
where
\begin{align*}
&U^{\rm P}_{1}\left(a\right)
={\rm Re}\int_{0}^{\pi/2}\frac{\mathrm{d}x}{B-a},
&&U^{\rm P}_{2}\left(a, b\right)
={\rm Re}\int_{0}^{\pi/2}\frac{\mathrm{d}x}{\left(B+b\right)^{2}+a^{2}},
\\
&U^{\rm P}_{3}\left(a, b\right)
={\rm Re}\int_{0}^{\pi/2}\frac{\left(B+b\right)\mathrm{d}x}{\left(B+b\right)^{2}+a^{2}},
&& U^{\rm P}_{4}
={\rm Re}\int_{0}^{\pi/2}1 \mathrm{d}x,
\\
&U^{\rm P}_{5}
={\rm Re}\int_{0}^{\pi/2}B \mathrm{d}x,
&& U^{\rm P}_{6}
={\rm Re}\int_{0}^{\pi/2}B^{2} \mathrm{d}x,
\end{align*}
and
\begin{align*}
&
V^{\rm P}_{1}\left(a\right)
={\rm Im}\int_{T_{\rm P}\cos{\theta}}^{T_{\rm up}^{\rm P}}
\frac{1}{B-a}\frac{\mathrm{d}B}{\sqrt{Q^{\rm P}_{2}(B)}},
&& 
V^{\rm P}_{2}\left(a, b\right)
={\rm Im}\int_{T_{\rm P}\cos{\theta}}^{T_{\rm up}^{\rm P}}
\frac{1}{\left(B+b\right)^{2}+a^{2}}
\frac{\mathrm{d}B}{\sqrt{Q^{\rm P}_{2}(B)}},
\\
&
V^{\rm P}_{3}\left(a, b\right)
={\rm Im}\int_{T_{\rm P}\cos{\theta}}^{T_{\rm up}^{\rm P}}
\frac{B+b}{\left(B+b\right)^{2}+a^{2}}
\frac{\mathrm{d}B}{\sqrt{Q^{\rm P}_{2}(B)}},
&& 
V^{\rm P}_{4}
={\rm Im}\int_{T_{\rm P}\cos{\theta}}^{T_{\rm up}^{\rm P}}
\frac{\mathrm{d}B}{\sqrt{Q^{\rm P}_{2}(B)}},
\\
&
V^{\rm P}_{5}
={\rm Im}\int_{T_{\rm P}\cos{\theta}}^{T_{\rm up}^{\rm P}}
\frac{B\mathrm{d}B}{\sqrt{Q^{\rm P}_{2}(B)}},
&& 
V^{\rm P}_{6}
={\rm Im}\int_{T_{\rm P}\cos{\theta}}^{T_{\rm up}^{\rm P}}
\frac{B^{2}\mathrm{d}B}{\sqrt{Q^{\rm P}_{2}(B)}}.
\end{align*}
We should pay particular attention to the functions $U^{\rm P}_{2}$, $U^{\rm P}_{3}$, $V^{\rm P}_{2}$ and $V^{\rm P}_{3}$ which contain $\sigma^{\rm P}_5$ or $\sigma^{\rm P}_6$. Since the properties of the Rayleigh wave only depends on these terms. 
The terms about the Rayleigh wave will be separated from the whole Green's function in the numerical examples to corroborate the argument.

In the next step, all the $U_i^{\rm P}\ (i=1,2,\cdots,6)$ will be calculated one by one.  
$U_{1}^{\rm P}\left(a\right)$ could be calculated by applying the method in Mooney ({\co 1974}), and the process to solve $U_{2}^{\rm P}\left(a, b\right)$ and $U_{3}^{\rm P}\left(a, b\right)$ is expounded in Appendix B of FZ18. The other integrations $U_4^{\rm P}$, $U_5^{\rm P}$ and $U_6^{\rm P}$ can be easily derived.
The ultimate results of $U_i^{\rm P}\ (i=1,2,\cdots,6)$ are displayed as followed:
\begin{align}
\label{UP12}
&U_{1}^{\rm P}\left(a\right)=\frac{\pi}{2}\frac{{\rm sgn}\left(T_{\rm P}\cos{\theta}-a\right)}{\sqrt{a^{2}-2aT_{\rm P}\cos{\theta}+T_{\rm P}^{2}-\sin^{2}{\theta}}},
&& U_{2}^{\rm P}\left(a, b\right)=\frac{\pi}{2a}{\rm Re}\left(y^{-1/2}\right),
\\
&U_{3}^{\rm P}\left(a, b\right)=-\frac{\pi}{2}{\rm Im}\left(y^{-1/2}\right),
&& U_{4}^{\rm P}=\frac{\pi}{2},
\\
\label{UP56}
&U_{5}^{\rm P}=\frac{\pi}{2}T_{\rm P}\cos{\theta},
&& U_{6}^{\rm P}=\frac{\pi}{2}T_{\rm P}^{2}\cos^{2}{\theta}-\frac{\pi}{4}(T_{\rm P}^{2}-1)\sin^{2}{\theta},
\end{align}
where
$y = \left(a+{\rm i}b+{\rm i}T_{\rm P}\cos\theta\right)^2
+\left(1-T_{\rm P}^2\right)\sin^2\theta$.

To express every $V_i^{\rm P}\ (i=1,2,\cdots,6)$, two real parameters $\xi_{1}$ and
$\xi_{2}$ ($<\xi_1$) have been introduced, which satisfy
\begin{align}
\label{P xi}\xi^{2}-\frac{T_{\rm P}^{2}+\cos^{2}{\theta}-k^{2}}{T_{\rm P}\cos{\theta}}\xi+1-k^{2}=0.
\end{align}
It is easy to show that $\xi_2<0<T_{\rm P}\cos\theta<\xi_1$. The complete procedures are rather tedious; however, the main steps have been outlined in FZ18. For simplicity only the final results of $V_i^{\rm P}\ (i=1,2,\cdots,6)$ are listed as:
\begin{align}
\notag
V_{1}^{\rm P}(a)
=&
-\left[H\left(a-T_{\rm P}\cos{\theta}\right)-H\left(a-\xi_{1}\right)\right]
\frac{1}{\sqrt{a^{2}+k^{2}-1}}
\frac{\pi}
{\sqrt{a^{2}-2aT_{\rm P}\cos{\theta}+T_{\rm P}^{2}-\sin^{2}{\theta}}} 
\\
\label{VP1}
&+\frac{1}{\xi_{2}-a}M_{\rm P}K_{\rm P}
+\frac{\xi_{2}-\xi_{1}}{\left(\xi_{1}-a\right)\left(\xi_{2}-a\right)}M_{\rm P}\Pi_{\rm P}
\left(-\left[
\frac{\left(\xi_{2}-a\right)C^{\rm P}_{2}}{\xi_{1}-a}
\right]^{2}\right),
\\
\notag
V^{\rm P}_{2}\left(a,b\right)
=&
\frac{M_{\rm P}}{a^{2}+\left(\xi_{2}+b\right)^{2}}
\left\{2\left(\xi_{1}-\xi_{2}\right)
{\rm Re}\left[\frac{m_{1}}{n_{1}}
\Pi_{\rm P}\left(-\left(C^{\rm P}_{2}\right)^{2}
\left(a^{2}+\left(\xi_{2}+b\right)^{2}\right)^{2}/n_{1}\right)
\right]
+K_{\rm P}\right\}
\\
\label{VP2}
&
-{\rm Im}\left[H\left(-b-T_{\rm P}\cos\theta\right)
H\left(-\left(b+\xi_1\right)\left(b+\xi_2\right)-a^2\right)
\frac{\pi}{a\sqrt{Q^{\rm P}_{2}(-b+{\rm i}a)}}\right],
\\
\notag
V^{\rm P}_{3}\left(a,b\right)
=&
\frac{M_{\rm P}}{a^{2}+\left(\xi_{2}+b\right)^{2}}
\left\{-2\left(\xi_{1}-\xi_{2}\right)a
{\rm Im}\left[
\frac{m_{1}}{n_{1}}\Pi_{\rm P}\left(-\left(C^{\rm P}_{2}\right)^{2}\left(a^{2}+\left(\xi_{2}+b\right)^{2}\right)^{2}/n_{1}\right)
\right]
+\left(\xi_{2}+b\right)K_{\rm P}
\right\}
\\
\label{VP3}
&
-{\rm Re}\left[ H\left(-b-T_{\rm P}\cos\theta\right)
H\left(-\left(b+\xi_1\right)\left(b+\xi_2\right)-a^2\right)
\frac{\pi}{\sqrt{Q^{\rm P}_{2}(-b+{\rm i}a)}}
\right],
\\
\label{VP4}
V_{4}^{\rm P}
=&
M_{\rm P}K_{\rm P},
\\
\label{VP5}
V_{5}^{\rm P}
=&
M_{\rm P}
\left[
\xi_{2}K_{\rm P}
+\left(\xi_{1}-\xi_{2}\right)\Pi_{\rm P}
\left(-\left(C_{2}^{\rm P}\right)^{2}\right)
\right],
\\
\label{VP6}
V_{6}^{\rm P}
=&
M_{\rm P}
\left[
T_{\rm P}\cos{\theta}\xi_{2}K_{\rm P}
+\xi_{1}\left(T_{\rm P}\cos{\theta}-\xi_{2}\right)
\left(E_{\rm P}-K_{\rm P}\right)
+T_{\rm P}\cos{\theta}\left(\xi_{1}-\xi_{2}\right)
\Pi_{\rm P}\left(-\left(C^{\rm P}_{2}\right)^{2}\right)
\right],
\end{align}
where
\begin{align*}
&
M_{\rm P}=\frac{1}{\sqrt{\xi_{1}(T_{\rm P}\cos{\theta}-\xi_{2})}}, 
&& 
\tau_{\rm P}=\sqrt{\frac{\xi_{2}(\xi_{1}-T_{\rm P}\cos{\theta})}{\xi_{1}(\xi_{2}-T_{\rm P}\cos{\theta})}}, 
&& 
C^{\rm P}_{2}=\sqrt{\frac{\xi_{1}-T_{\rm P}\cos{\theta}}{T_{\rm P}\cos{\theta}-\xi_{2}}},
\\
&
K_{\rm P}=K(\tau_{\rm P}), 
&&
E_{\rm P}=E(\tau_{\rm P}), 
&& 
\Pi_{\rm P}(\cdot)=\Pi(\tau_{\rm P}, \cdot),
\end{align*}
and
\begin{align*}
m_{1}=&
\frac{1}{2}\left(\xi_{1}+\xi_{2}+2b\right)
\left(a^{2}+\left(\xi_{2}+b\right)^{2}\right)
-2\left(\xi_{2}+b\right)
\left(a^{2}+\left(\xi_{1}+b\right)\left(\xi_{2}+b\right)\right)
\\
&
-{\rm i}
\left[
\frac{a^{2}+\left(\xi_{1}+b\right)\left(\xi_{2}+b\right)}{2a}
\left(a^{2}-\left(\xi_{2}+b\right)^{2}\right)
+\left(\xi_{2}+b\right)\left(\xi_{1}-\xi_{2}\right)a
\right],
\\
n_{1}=&
2\left(a^{2}+\left(\xi_{1}+b\right)
\left(\xi_{2}+b\right)\right)^{2}
-\left(a^{2}+\left(\xi_{1}+b\right)^{2}\right)
\left(a^{2}+\left(\xi_{2}+b\right)^{2}\right)
\\
&
+2{\rm i}a
\left(a^{2}+\left(\xi_{1}+b\right)\left(\xi_{2}+b\right)\right)\left(\xi_{1}-\xi_{2}\right).
\end{align*}
The expressions of $U^{\rm P}_i\ (i=1,4,5,6)$ and $V^{\rm P}_i\ (i=1,4,5,6)$ agree perfectly with those in FZ18, and the expressions of $U^{\rm P}_i\ (i=2,3)$ and $V^{\rm P}_i\ (i=2,3)$ in the two papers are quite similar. 

So far, the closed-form of $u_j^{\rm P}\ (j=1,2,3)$ are successfully derived.
The procedures are so elaborate that an explicit guide to our formulae should be listed below to allow readers to implement our results easily:
\\ \hspace*{\fill} \\
\begin{tikzpicture}
    [L2Node/.style={rectangle, draw=blue!30, fill=blue!10, very thick, minimum size=10mm},
    L1Node/.style={rectangle, draw=black!70, fill=gray!15, very thick, minimum size=10mm},
    L3Node/.style={rectangle, draw=green!50,fill=green!10,very thick, minimum size=10mm}]
\node[L1Node] (original formular) at (-2, 3) {Original Integral};
\node[L1Node] (standard formular) at (3, 3) {Standard Integral};
\node[L1Node] (fraction1) at (6, 4.5) {Elementary Coefficients};
\node[L1Node] (fraction2) at (6, 1.5) {Elliptic Coefficients};
\draw [black, thick][-latex] (original formular)->(standard formular);
\coordinate (A) at (3, 4.5) coordinate (B) at (3, 1.5) coordinate (C) at (8.5, 4.5) coordinate (D) at (8, 1.5) coordinate (E) at (10, 4.5) coordinate (F) at (10, 1.5);
\draw [black, thick][-latex] (standard formular)--(A)--(fraction1);
\draw [black, thick][-latex] (standard formular)--(B)--(fraction2);
\node[L1Node] (final formular) at (9, 3) {Final Solution};
\draw [black, thick][-latex] (fraction1)--(C)--(final formular);
\draw [black, thick][-latex] (fraction2)--(D)--(final formular);
\node[L1Node] (U) at (12, 4.5) {Elementary Functions};
\node[L1Node] (V) at (12, 1.5) {Elliptic Functions};
\draw [black, thick][->] (U)--(E)--(final formular);
\draw [black, thick][->] (V)--(F)--(final formular);
\end{tikzpicture}
\begin{enumerate}
\item Original Integral (eq. (\ref{Bakker uPc})): Integral solutions in BVK99 after our reorganization.
\item Standard Integral (eq. (\ref{standard uPc})): With a substitution $B = \alpha\eta_{\alpha}$ to convert the form of original integral to an equivalent form, which contains the elementary integral and the elliptic integral.
\item Elementary Coefficients and Elliptic Coefficients 
(eqs. (\ref{elementary fraction}) and (\ref{elliptic fraction})): 
determined by the partial fraction expansions. The Matlab command ``residue'' could output the same results directly.
\item Elementary Functions (eqs. (\ref{UP12})--(\ref{UP56})): 
explicit expressions of $U^{\rm P}_i\ (i=1,\cdots,6)$ which only contain elementary functions.  
\item Elliptic Functions (eqs. (\ref{VP1})--(\ref{VP6})): 
explicit expressions of $V^{\rm P}_i\ (i=1,\cdots,6)$ which contain three kinds of elliptic integrals as well as elementary functions.
\item Final Solution (eq. (\ref{final uPc})): 
Linear combine the paired coefficients and functions to derive the final solution regarded as closed-form.  
\end{enumerate}
\section{Calculation of $u_{\lowercase{j}}^{\rm S\lowercase{c}}$}
With the substitution $B=\beta\eta_{\beta}$, the original form of $u_j^{\rm Sc}$ (eq. (\ref{Bakker uSc})) are converted into the standard form as
\begin{align}
u_j^{\rm Sc}&
=\frac{1}{\pi^2\mu r}
H\left(T_{\rm S}-1\right)
\int_{T_{\rm S}\cos\theta}^{T^{\rm S}_{\rm up}}
{\rm Im}\Big[
\frac{M^{\rm S}_j\left(B\right)}{\sigma^{\rm S}\left(B\right)W^{\rm S}\left(B\right)}
\frac{{\rm d}B}{\sqrt{Q^{\rm S}_1\left(B\right)}}
+\frac{N^{\rm S}_j(B)}{\sigma^{\rm S}\left(B\right)W^{\rm S}\left(B\right)}
\frac{{\rm d}B}{\sqrt{Q^{\rm S}_2\left(B\right)}}
\Big],
\end{align}
where
\begin{align}\label{uSc parameter 1}
\notag
M^{\rm S}_j(B)&=8B^2\left(B^2-1\right)\left(B^2+k^{-2}-1\right) S_j\left(B\right),\\
\notag
N^{\rm S}_j(B)&=2B\left(B^2+k^{-2}-1\right)\left(2B^2-1\right)^2 S_j\left(B\right),\\
S_1(B)&=-\tfrac{x_1^2}{r^2}B\left(k_{\beta} B^2-\tfrac{x_3}{x_1} B
+T_{\rm S}\tfrac{r}{x_1}-k_{\beta}\right)
+k_{\beta}B\left(B^2-2T_{\rm S}B\cos\theta+T_{\rm S}^2-\sin^2\theta\right),
\\
\notag
S_2(B)&=-\tfrac{x_1 x_2}{r^2}B\left(k_{\beta} B^2-\tfrac{x_3}{x_1} B+T_{\rm S}\tfrac{r}{x_1}-k_{\beta}\right),
\\
\notag
S_3(B)&=-\tfrac{x_1}{r}\left(B^2-1\right)\left(k_{\beta}B^2\cos\theta
+\tfrac{R}{x_1}B\sin\theta-k_{\beta} T_{\rm S}B\right),
\end{align}
and
\begin{align}\label{uSc parameter 2}
\notag
&T^{\rm S}_{\rm up}
=T_{\rm S}\cos{\theta}+{\rm i}\sqrt{T_{\rm S}^{2}-1}\sin{\theta}, 
\\
&\sigma^{\rm S}(B)
=(2B^{2}-1)^{4}-16B^{2}(B^{2}+k^{-2}-1)(B^{2}-1)^{2},
\\
\notag
&W^{\rm S}(B)
=\left(k_{\beta}B\cos\phi \cos\theta+\sin\theta
-k_{\beta}T_{\rm S}\cos\phi\right)^2+k^2_{\beta}\sin^2\phi
\left(B^2-2T_{\rm S}B\cos\theta+T^2_{\rm S}-\sin^2\theta\right),
\\
\notag
&Q^{\rm S}_1(B)
=B^{2}-2T_{\rm S}B\cos{\theta}+T_{\rm S}^{2}-\sin^{2}{\theta}, 
\\
\notag
&Q^{\rm S}_2(B)
=(B^{2}+k^{-2}-1)(B^{2}-2T_{\rm S}B\cos{\theta}+T_{\rm S}^{2}-\sin^{2}{\theta}),
\end{align}
The closed-form solution for $u^{\rm Sc}_{j}$ are obtained in a way highly similar to those for $u^{\rm Pc}_{j}$, so only the difference of the two results will be pointed out. 
The final formula of $u^{\rm Sc}_{j}$ could be written as
\begin{align}\label{final uSc}
\notag
u^{\rm Sc}_{j}=&\frac{1}{\pi^{2}\mu r}
H\left(T_{\rm S}-1\right)
\Big[
\sum_{s=1}^4 u^{\rm S}_{j,s}U^{\rm S}_{1}\left(\sigma^{\rm S}_s\right)
+u^{\rm S}_{j,5}U^{\rm S}_{2}\left({\rm Im}\left(\sigma^{\rm S}_{5}\right), 0\right)
+u^{\rm S}_{j,6}U^{\rm S}_{3}\left({\rm Im}\left(\sigma^{\rm S}_{6}\right), 0\right)
\\&
\notag
+u^{\rm S}_{j,7}U^{\rm S}_{2}\left({\rm Im}\left(\omega^{\rm S}_{1}\right), -{\rm Re}\left(\omega^{\rm S}_{1}\right)\right)
+u^{\rm S}_{j,8}U^{\rm S}_{3}\left({\rm Im}\left(\omega^{\rm S}_{2}\right), -{\rm Re}\left(\omega^{\rm S}_{2}\right)\right)
+u^{\rm S}_{j,9}U^{\rm S}_{4}+u^{\rm S}_{j,10}U^{\rm S}_{5}
\Big]
\\&
\notag
+\frac{1}{\pi^{2}\mu r}
H\left(T_{\rm S}-1\right)
\Big[\sum_{s=1}^4 v^{\rm S}_{j,s}V^{\rm P}_{1}\left(\sigma^{\rm S}_s\right)
+v^{\rm S}_{j,5}V^{\rm S}_{2}\left({\rm Im}\left(\sigma^{\rm S}_{5}\right), 0\right)
+v^{\rm S}_{j,6}V^{\rm S}_{3}\left({\rm Im}\left(\sigma^{\rm S}_{6}\right), 0\right)
\\&
+v^{\rm S}_{j,7}V^{\rm S}_{2}\left({\rm Im}\left(\omega^{\rm S}_{1}\right), -{\rm Re}\left(\omega^{\rm S}_{1}\right)\right)
+v^{\rm S}_{j,8}V^{\rm S}_{3}\left({\rm Im}\left(\omega^{\rm S}_{2}\right), -{\rm Re}\left(\omega^{\rm S}_{2}\right)\right)
+v^{\rm S}_{j,9}V^{\rm S}_{4}
+v^{\rm S}_{j,10}V^{\rm P}_{S}
+v^{\rm S}_{j,11}V^{\rm S}_{6}\Big],
\end{align}
where
\begin{align}
\notag
V^{\rm S}_{1}(a)
=&\frac{M_{\rm S}}{\xi_{2}-a}
\left[
\frac{(\xi_{1}-a)(\xi_{2}-T_{\rm S}\cos{\theta})}
{a^{2}+T_{\rm S}^{2}-2aT_{\rm S}\cos{\theta}-\sin^{2}{\theta}}
\Pi_{\rm S}
\left(
\frac{\left(C^{\rm S}_{2}\right)^{2}\left(\xi_{2}-a\right)^{2}}{\left(C^{\rm S}_{2}\right)^{2}\left(\xi_{2}-a\right)^{2}+\left(\xi_{1}-a\right)^{2}}
\right)
+K_{\rm S}
\right]
\\
\label{VS1}
&
-\left(H\left(a-T_{\rm S}\cos{\theta}\right)
-H\left(a-\xi_{1}\right)\right)
\frac{\pi}{\sqrt{a^{2}+k^{-2}-1}
\sqrt{a^{2}-2aT_{\rm S}\cos{\theta}+T_{\rm S}^{2}-\sin^{2}{\theta}}},
\\
\notag
V^{\rm S}_{2}\left(a,b\right)
=&
\frac{M_{\rm S}}{a^{2}+\left(\xi_{2}+b\right)^{2}}
\left\{
2\left(\xi_{1}-\xi_{2}\right)
{\rm Re}\left[
\frac{m_{1}}{n_{1}}
\Pi_{\rm S}\left(\left(C^{\rm S}_{2}\right)^{2}\left(a^{2}+\left(\xi_{2}+b\right)^{2}\right)^{2}/n_{1}\right)
\right]
+K_{\rm S}\right\}
\\
\label{VS2}
&
-{\rm Im}\left[
H\left(-b-T_{\rm S}\cos\theta\right)
H\left(-\left(b+\xi_1\right)\left(b+\xi_2\right)-a^2\right)
\frac{\pi}{a\sqrt{Q^{\rm S}_{2}(-b+{\rm i}a)}}
\right],
\\
\notag
V^{\rm S}_{3}\left(a,b\right)
=&
\frac{M_{\rm S}}{a^{2}+\left(\xi_{2}+b\right)^{2}}
\left\{
-2\left(\xi_{1}-\xi_{2}\right)a
{\rm Im}\left[
\frac{m_{1}}{n_{1}}
\Pi_{\rm S}\left(\left(C^{\rm S}_{2}\right)^{2}\left(a^{2}+\left(\xi_{2}+b\right)^{2}\right)^{2}/n_{1}\right)
\right]
+\left(\xi_{2}+b\right)K_{\rm S}
\right\}
\\
\label{VS3}
&
-{\rm Re}\left[ 
H\left(-b-T_{\rm S}\cos\theta\right)
H\left(-\left(b+\xi_1\right)\left(b+\xi_2\right)-a^2\right)
\frac{\pi}{\sqrt{Q^{\rm S}_{2}(-b+{\rm i}a)}}
\right],
\\
\label{VS4}
V^{\rm S}_{4}
=&M_{\rm S}K_{\rm S},
\\
\label{VS5}
V^{\rm S}_{5}
=&M_{\rm S}
\left[
\xi_{2}K_{\rm S}
+\left(T_{\rm S}\cos{\theta}-\xi_{2}\right)
\Pi_{\rm S}\left(\frac{\xi_{1}-T_{\rm S}\cos{\theta}}{\xi_{1}-\xi_{2}}\right)
\right],
\\
\notag
V^{\rm S}_{6}
=&M_{\rm S}
\Big[
T_{\rm S}\cos{\theta}\left(\xi_{1}-\xi_{2}\right)E_{\rm S}
+\left(\xi_{1}\xi_{2}-\left(\xi_{1}-\xi_{2}\right)T_{\rm S}\cos{\theta}\right)K_{\rm S}
\\
\label{VS6}
&
+\left(T_{\rm S}\cos{\theta}-\xi_{2}\right)T_{\rm S}\cos{\theta}
\Pi_{\rm S}\left(\frac{\xi_{1}-T_{\rm S}\cos{\theta}}{\xi_{1}-\xi_{2}}\right)
\Big],
\end{align}
where
\begin{align*}
&C^{\rm S}_{2}
=\sqrt{\frac{\xi_{1}-T_{\rm S}\cos{\theta}}{T_{\rm S}\cos{\theta}-\xi_{2}}}, 
&& M_{\rm S}
=\frac{1}{\sqrt{T_{\rm S}\cos{\theta}(\xi_{1}-\xi_{2})}}, 
&&\tau_{\rm S}
=\sqrt{\frac{\xi_{2}(\xi_{1}-T\cos{\theta})}
{T_{\rm S}\cos{\theta}(\xi_{1}-\xi_{2})}}, 
\\
&K_{\rm S}
=K(\tau_{\rm S}), 
&& E_{\rm S}
=E(\tau_{\rm S}), 
&&\Pi_{\rm S}(\cdot) 
=\Pi(\tau_{\rm S},\cdot),
\end{align*}
and
\begin{align*}
m_{1}=&
\frac{1}{2}\left(\xi_{1}+\xi_{2}+2b\right)
\left(a^{2}+\left(\xi_{2}+b\right)^{2}\right)
-2\left(\xi_{2}+b\right)\left(a^{2}
+\left(\xi_{1}+b\right)\left(\xi_{2}+b\right)\right)
\\&
-{\rm i}\left[
\frac{a^{2}+\left(\xi_{1}+b\right)\left(\xi_{2}+b\right)}{2a}
\left(a^{2}-\left(\xi_{2}+b\right)^{2}\right)
+\left(\xi_{2}+b\right)\left(\xi_{1}-\xi_{2}\right)a
\right],
\\
n_{1}=&
\left(C^{\rm S}_2\right)^2
\left(a^2+\left(\xi_2+b\right)^2\right)^2
+2\left(a^{2}+\left(\xi_{1}+b\right)\left(\xi_{2}+b\right)\right)^{2}
-\left(a^{2}+\left(\xi_{1}+b\right)^{2}\right)
\left(a^{2}+\left(\xi_{2}+b\right)^{2}\right)
\\&
+2{\rm i}a\left(a^{2}+\left(\xi_{1}+b\right)
\left(\xi_{2}+b\right)\right)\left(\xi_{1}-\xi_{2}\right).
\end{align*}
All the coefficients of equation (\ref{final uSc}) are determined by $\frac{M^{\rm S}_j\left(B\right)}{\sigma^{\rm S}\left(B\right)W^{\rm S}(B)}$ and $\frac{N^{\rm S}_j\left(B\right)}{\sigma^{\rm S}\left(B\right)W^{\rm S}(B)}$ as: 
\begin{align}
\notag
\frac{M_j\left(B\right)}{\sigma^{\rm S}(B)W^{\rm S}(B)}
=&\sum_{s=1}^{4}\frac{u^{\rm S}_{j,s}}{B-\sigma^{\rm S}_{s}}
+\frac{u^{\rm S}_{j,5}}{B^2+\left({\rm Im}\left(\sigma^{\rm S}_{5}\right)\right)^2}
+\frac{u^{\rm S}_{j,6}B}{B^2+\left({\rm Im}\left(\sigma^{\rm S}_{6}\right)\right)^2}
+\frac{u^{\rm S}_{j,7}}{\left(B-{\rm Re}\left(\omega^{\rm S}_1\right)\right)^2
+\left({\rm Im}\left(\omega^{\rm S}_{1}\right)\right)^2}
\\&
\label{elementary fraction S}
+\frac{u^{\rm S}_{j,8}B}{\left(B-{\rm Re}\left(\omega^{\rm S}_2\right)\right)^2
+\left({\rm Im}\left(\omega^{\rm S}_{2}\right)\right)^2}
+u^{\rm S}_{j,9}+u^{\rm S}_{j,10}B,
\\
\notag
\frac{N_j\left(B\right)}{\sigma^{\rm S}(B)W^{\rm S}(B)}
=&\sum_{s=1}^{4}\frac{v^{\rm S}_{j,s}}{B-\sigma^{\rm S}_{s}}
+\frac{v^{\rm S}_{j,5}}{B^2+\left({\rm Im}\left(\sigma^{\rm S}_{5}\right)\right)^2}
+\frac{v^{\rm S}_{j,6}B}{B^2+\left({\rm Im}\left(\sigma^{\rm S}_{6}\right)\right)^2}
+\frac{v^{\rm S}_{j,7}}{\left(B-{\rm Re}\left(\omega^{\rm S}_1\right)\right)^2
+\left({\rm Im}\left(\omega^{\rm P}_{1}\right)\right)^2}
\\&
\label{elliptic fraction S}
+\frac{v^{\rm S}_{j,8}B}{\left(B-{\rm Re}\left(\omega^{\rm S}_2\right)\right)^2
+\left({\rm Im}\left(\omega^{\rm S}_{2}\right)\right)^2}
+v^{\rm S}_{j,9}+v^{\rm S}_{j,10}B+v^{\rm S}_{j,11}B^2.
\end{align}
Two purely imaginary roots of the polynomial $\sigma^{\rm S}\left(B\right)$ are designated as $\sigma^{\rm S}_5$ and $\sigma^{\rm S}_6$, which involve the properties of the Rayleigh wave.
The results of $U^{\rm S}_{i}\ (i=1,2,\cdots,6)$ are left out since they are identical to $U^{\rm P}_{i}\ (i=1,2,\cdots,6)$, except changing $T_{\rm P}$ to $T_{\rm S}$. In addition, the equation to determine $\xi_1$ and $\xi_2$ ($\xi_1>\xi_2$) is changed to
\begin{align}
\label{S wave xi}\xi^{2}-\frac{T_{\rm S}^{2}+\cos^{2}{\theta}-k^{-2}}{T_{\rm S}\cos{\theta}}\xi+1-k^{-2}=0.
\end{align}
\section{Calculation of $u_{\lowercase{j}}^{\rm S\text{-}P\lowercase{c}}$}
Following the same method of solving the integration of $u^{\rm Sc}_{j}$,
the closed-form formula of $u^{\rm S\text{-}Pc}_{j}$ could be written as
\begin{align}
\notag
u^{\rm S\text{-}Pc}_{j}=&
\frac{1}{\pi^{2}\mu r}
H\left(T_{\rm S}-1\right)
\Big[\sum_{s=1}^4 v^{\rm S}_{j,s}V^{\rm S\text{-}P}_{1}\left(\sigma^{\rm S}_s\right)
+v^{\rm S}_{j,5}V^{\rm S\text{-}P}_{2}\left({\rm Im}\left(\sigma^{\rm S}_{5}\right), 0\right)
+v^{\rm S}_{j,6}V^{\rm S\text{-}P}_{3}\left({\rm Im}\left(\sigma^{\rm S}_{6}\right), 0\right)
\\&
\notag
+v^{\rm S}_{j,7}V^{\rm S\text{-}P}_{2}\left({\rm Im}\left(\omega^{\rm S}_{1}\right), -{\rm Re}\left(\omega^{\rm S}_{1}\right)\right)
+v^{\rm S}_{j,8}V^{\rm S\text{-}P}_{3}\left({\rm Im}\left(\omega^{\rm S}_{2}\right), -{\rm Re}\left(\omega^{\rm S}_{2}\right)\right)
\\&
\label{final uSPc}
+v^{\rm S}_{j,9}V^{\rm S\text{-}P}_{4}
+v^{\rm S}_{j,10}V^{\rm S\text{-}P}_{5}
+v^{\rm S}_{j,11}V^{\rm S\text{-}P}_{6}\Big],
\end{align}
where
\begin{align}
\notag
V^{\rm S\text{-}P}_{1}\left(a\right)
=&\frac{M_{\rm S\text{-}P}}{\xi_{2}-a}
\left[
\frac{\left(\xi_{2}-T_{\rm S}\cos{\theta}\right)\left(\xi_{1}-a\right)}{a^{2}-2aT_{\rm S}\cos{\theta}+T_{\rm S}^{2}-\sin^{2}{\theta}}
\Pi_{\rm S\text{-}P}\left(c\right)+K_{\rm S\text{-}P}
\right]
\\
\label{VSP1}
&-\frac{\pi}
{2\sqrt{\left(a^{2}-1+k^{-2}\right)
\left(a^{2}-2aT_{\rm S}\cos{\theta}+T_{\rm S}^{2}-\sin^{2}{\theta}\right)}},
\\
\notag
V^{\rm S\text{-}P}_{2}\left(a, b\right)=&
\frac{M_{\rm S\text{-}P}K_{\rm S\text{-}P}}{a^{2}+\left(\xi_{2}+b\right)^{2}}
+M_{\rm S\text{-}P}\left(T_{\rm S}\cos{\theta}-\xi_{2}\right)a^{-1}
{\rm Im}
\left[
\frac{a+{\rm i}\left(\xi_{1}+b\right)}{a+{\rm i}\left(\xi_{2}+b\right)}
y_1^{-1}\Pi_{\rm S\text{-}P}\left(-c_{1}\right)
\right]
\\
\label{VSP2}
&
+\frac{\pi}{2a}{\rm Im}
\left[y_1^{-1/2}\left(\left(a+{\rm i}b\right)^{2}+1-k^{-2}\right)^{-1/2}\right],
\\
\notag
V^{\rm S\text{-}P}_{3}\left(a, b\right)=&
\frac{M_{\rm S\text{-}P}K_{\rm S\text{-}P}\left(\xi_2+b\right)}{a^{2}+\left(\xi_{2}+b\right)^{2}}
+M_{\rm S\text{-}P}\left(T_{\rm S}\cos{\theta}-\xi_{2}\right)
{\rm Re}
\left[
\frac{a+{\rm i}\left(\xi_{1}+b\right)}{a+{\rm i}\left(\xi_{2}+b\right)}
y_1^{-1}\Pi_{\rm S\text{-}P}\left(-c_{1}\right)
\right]
\\
\label{VSP3}
&
+\frac{\pi}{2}{\rm Re}
\left[y_1^{-1/2}\left(\left(a+{\rm i}b\right)^{2}+1-k^{-2}\right)^{-1/2}\right],
\\
\label{VSP4}
V^{\rm S\text{-}P}_{4}
=&M_{\rm S\text{-}P}K_{\rm S\text{-}P},
\\
\label{VSP5}
V^{\rm S\text{-}P}_{5}
=&M_{\rm S\text{-}P}\left[
\xi_{2}K_{\rm S\text{-}P}
+\left(T_{\rm S}\cos{\theta}-\xi_{2}\right)\Pi_{\rm S\text{-}P}\left(T_{\rm S}\cos{\theta}/\xi_{2}\right)
\right]
+\frac{\pi}{2},
\\
\notag
V^{\rm S\text{-}P}_{6}
=&M_{\rm S\text{-}P}\Big[
\xi_{2}T_{\rm S}\cos{\theta}K_{\rm S\text{-}P}
+\xi_{2}\left(\xi_{1}-T_{\rm S}\cos{\theta}\right)E_{\rm S\text{-}P}
\\&
\label{VSP6}
+\left(T_{\rm S}\cos{\theta}-\xi_{2}\right)T_{\rm S}\cos{\theta}\Pi_{\rm S\text{-}P}\left(T_{\rm S}\cos{\theta}/\xi_{2}\right)
\Big]
+\frac{\pi}{2}T_{\rm S}\cos{\theta},
\end{align}
where
\begin{align*}
&M_{\rm S\text{-}P}=\frac{1}{\sqrt{\xi_{2}\left(\xi_{1}-T_{\rm S}\cos{\theta}\right)}}, &&
\tau_{\rm S\text{-}P}=\sqrt{\frac{T_{\rm S}\cos{\theta}\left(\xi_{1}-\xi_{2}\right)}{\xi_{2}\left(\xi_{1}-T_{\rm S}\cos{\theta}\right)}}, &&
c=\frac{T_{\rm S}\cos{\theta}\left(\xi_{2}-a\right)^{2}}{\left(a^{2}-2aT_{\rm S}
\cos{\theta}+T_{\rm S}^{2}-\sin^{2}{\theta}\right)\xi_{2}},\\
&K\left(\tau_{\rm S\text{-}P}\right)=K_{\rm S\text{-}P}, && E\left(\tau_{\rm S\text{-}P}\right)=E_{\rm S\text{-}P}, && \Pi\left(\tau_{\rm S\text{-}P}, \cdot\right)=\Pi_{\rm S\text{-}P}\left(\cdot\right),
\end{align*}
and
\begin{align*}
&c_{1}=\frac{T_{\rm S}\cos{\theta}\left(\xi_{2}-{\rm i}a\right)^{2}}{\left(\sin^{2}{\theta}+a^{2}-T_{\rm S}^2+2{\rm i}aT\cos{\theta}\right)\xi_{2}},
&&y_1=\sin^{2}{\theta}+a^{2}-T_{\rm S}^2+2{\rm i}aT_{\rm S}\cos{\theta}.
\end{align*}
\section{Ultimate results for $u_j$}
Up to now, the original integrations of $u^{\rm Pc}_j$, $u^{\rm Sc}_j$ and $u^{\rm S\text{-}Pc}_j$ have been decomposed into the corresponding closed-form  expressions as eq. (\ref{final uPc}), eq. (\ref{final uSc}) and eq. (\ref{final uSPc}), respectively. 
Therefore, the complete closed-form formulae for the whole displacements $u_j$ have been obtained. However, the solution scheme is sufficiently complicated that readers may not know what the ultimate formulae are.
A summarization for readers should be listed to make our results used easily: 
\begin{enumerate}
\item $u_j$ is calculated by (\ref{Full expression}), which consists of six parts: 
$u^{\rm Pc}_j$, $u^{\rm Sc}_j$, $u^{\rm S\text{-}Pc}_j$, 
 $u^{\rm Pl}_j$, $u^{\rm Sl}_j$  and $u^{\rm S\text{-}Pl}_j$.
\item $u^{\rm Pl}_j$, $u^{\rm Sl}_j$ and $u^{\rm S\text{-}Pl}_j$ are calculated by eqs. (\ref{Bakker uPl}), (\ref{Bakker uSl}) and (\ref{Bakker uSPl}), respectively.
\item $u^{\rm Pc}_j$ is calculated by (\ref{final uPc}), which consists of the functions labeled as $U^{\rm P}$ and $V^{\rm P}$, and the coefficents labeled as $u^{\rm P}$, $v^{\rm P}$, $\sigma^{\rm P}$ and 
$\omega^{\rm P}$:
\begin{enumerate}
\item The functions in (\ref{final uPc}) are calculated by (\ref{UP12})--(\ref{UP56}) and (\ref{VP1})--(\ref{VP6}).
\item The coefficents in (\ref{final uPc}) are calculated by (\ref{elementary fraction}) and (\ref{elliptic fraction}), and related terms are in (\ref{uPc parameter1}) and (\ref{uPc parameter2}).
\end{enumerate}
\item $u^{\rm Sc}_j$ is calculated by (\ref{final uSc}), which consists of the functions labeled as $U^{\rm S}$ and $V^{\rm S}$, and the coefficents labeled as $u^{\rm S}$, $v^{\rm S}$, $\sigma^{\rm S}$ and 
$\omega^{\rm S}$:
\begin{enumerate}
\item The functions in (\ref{final uSc}) are calculated by (\ref{VS1})--(\ref{VS6}).
\item The coefficents in (\ref{final uSc}) are calculated by (\ref{elementary fraction S}) and (\ref{elliptic fraction S}), and related terms are in (\ref{uSc parameter 1}) and (\ref{uSc parameter 2}).
\end{enumerate}
\item $u^{\rm S\text{-}Pc}_j$ is calculated by (\ref{final uSPc}), which consists of the functions labeled as $V^{\rm S\text{-}P}$, and the coefficents labeled as $v^{\rm S}$, $\sigma^{\rm S}$ and 
$\omega^{\rm S}$:
\begin{enumerate}
\item The functions in (\ref{final uSPc}) are calculated by (\ref{VSP1})--(\ref{VSP6}).
\item The coefficents in (\ref{final uSPc}) are the same to those in 
$u^{\rm Sc}_j$.
\end{enumerate}
\end{enumerate}
Furthermore, a Matlab program for implementing our solution will be provided as Supporting Information on the journal site, to allow readers to reproduce our results.
\section{Numerical Results}
In this section, a few examples of computed Green's functions are compared with the reorganized results of BVK99's to verify the correctness of the formulae obtained in previous sections. 
\begin{figure}
\centering
\includegraphics[width=.6\textwidth]{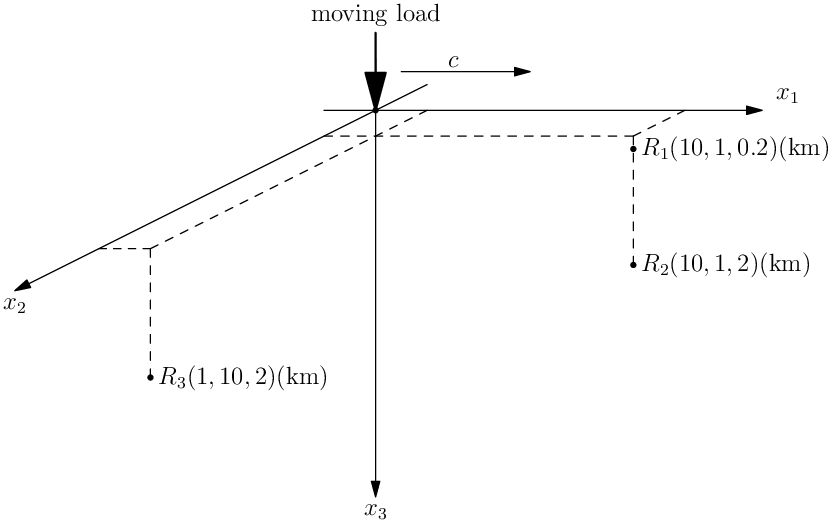}
\caption{The geometry of the numerical instances. $R_1$ denotes the receivers in Fig. 3 and Fig. 7, $R_2$ in Figs. 4--6 and $R_3$ in Fig. 8, respectively.}
\end{figure}
In all the following numerical calculations, the material parameters are set as: $\alpha=8.00\ \rm{km/s}$, $\beta=4.62\ {\rm km/s}$, and $\rho=3.30\ {\rm g/cm^3}$, the same as those in Johnson ({\co 1974}).
Fig. 2 shows the coordinates of the receivers in the numerical instances.
The receiver $R_1$ in Figs. 3 and 7 are located at ($10, 1, 2$) (km). $R_2$ in Figs. 4-6 and $R_3$ in Fig. 8 are at ($10, 1, 0.2$) (km) and ($1, 10, 2)$ (km), respectively.
\begin{figure}
\label{compareJohnson}
\centering
\subfigure{
\begin{minipage}{.48\textwidth}
\centering
\includegraphics[scale=0.42]
{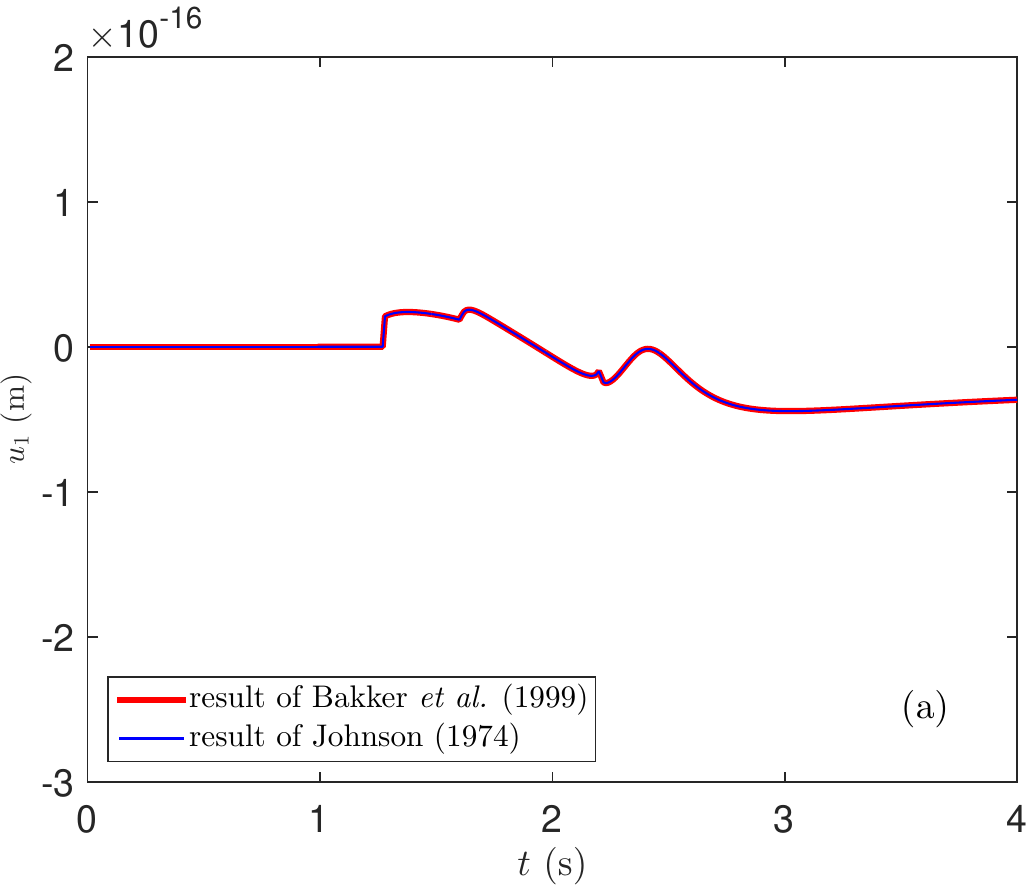}
\end{minipage}
}
\subfigure{
\begin{minipage}{.48\textwidth}
\centering
\includegraphics[scale=0.42]
{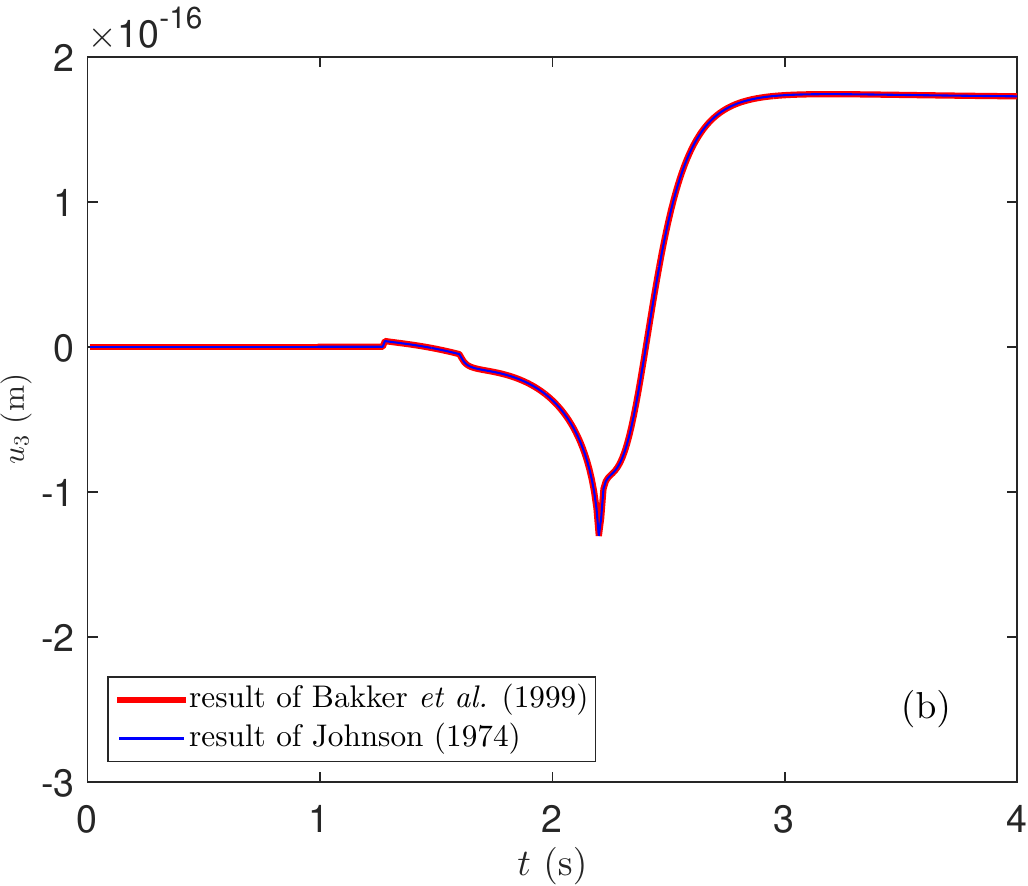}
\end{minipage}
}
\caption{(a) Comparison of displacements between 
Johnson's ({\color{blue}1974}) solutions and those in BVK99 after our reorganization, when velocity comes to zero. 
The receiver is at the $(10, 1, 2)$ (km).
(a) $u_{1}$; (b) $u_{3}$.}
\end{figure}

When $c$ decreases to zero, the results degenerate into the solutions of Lamb's problem (FZ18).   
In Fig. 3, displacement components $u_1$ and $u_3$ are shown to demonstrate the connection between the two classic problems, and the agreement of the wave fields verifies the correctness of the rewritten expressions. 

The correctness test on our results should be implemented by comparison with the results in BVK99. The comparisons for three components of the displacements $u_j$ are shown in Figs. 4 (a), 5 (a) and 6 (a). 
The velocity of the source $c$ is $2.00$ km/s. 
All of our results are identical to the reorganized solutions of BVK99's, which is sufficient to validate the formulae given in previous sections.

To analyze the behavior of the Rayleigh wave, $u_{j}$ is separated into two parts:
\begin{align}
u_{j}=RAY_{j}+O_{j},
\end{align}
where
\begin{small}
\begin{align*}
RAY_{j}=&
\frac{1}{\pi^{2}\mu r}
H\left(T_{\rm P}-1\right)
\left[
u^{\rm P}_{j,5}U^{\rm P}_{2}\left({\rm Im}\sigma^{\rm P}_{5}, 0\right)
+u^{\rm P}_{j,6}U^{\rm P}_{3}\left({\rm Im}\sigma^{\rm P}_{6}, 0\right)
+v^{\rm P}_{j,5}V^{\rm P}_{2}\left({\rm Im}\sigma^{\rm P}_{5}, 0\right)
+v^{\rm P}_{j,6}V^{\rm P}_{3}\left({\rm Im}\sigma^{\rm P}_{6}, 0\right)
\right]
\\&
+\frac{1}{\pi^{2}\mu r}
H\left(T_{\rm S}-1\right)
\left[
u^{\rm S}_{j,5}U^{\rm S}_{2}\left({\rm Im}\sigma^{\rm S}_{5}, 0\right)
+u^{\rm S}_{j,6}U^{\rm S}_{3}\left({\rm Im}\sigma^{\rm S}_{6}, 0\right)
+v^{\rm S}_{j,5}V^{\rm S}_{2}\left({\rm Im}\sigma^{\rm S}_{5}, 0\right)
+v^{\rm S}_{j,6}V^{\rm S}_{3}\left({\rm Im}\sigma^{\rm S}_{6}, 0\right)
\right],
\end{align*}
\end{small}
and $O_j=u_j-RAY_j$. $RAY_{j}$ is defined as the Rayleigh term, because it contains all the terms which involve the zero of the Rayleigh function $\sigma^{\rm P}_{5}$, $\sigma^{\rm P}_{6}$, $\sigma^{\rm S}_{5}$ and $\sigma^{\rm S}_{6}$. Correspondingly, $O_{j}$ is referred to as the ``Other'' term. For ``Other'' term, in Figs. 4(b), 5(b) and 6(b), the displacements always decay smoothly after the arrival time of S wave. For Rayleigh term in Figs. 4(b), 5(b) and 6(b), however, the Rayleigh wave becomes the dominant phase as the proportion of the depth of the receiver to $r$ (see Fig. 1 for reference) is small.
The numerical analysis verifies that the Rayleigh wave is generated by $RAY_{j}$, and we shall report on additional features of the Rayleigh wave in subsequent articles on this issue. 
\begin{figure}
\centering
\subfigure{
\begin{minipage}{.48\textwidth}
\centering
\includegraphics[scale=0.42]
{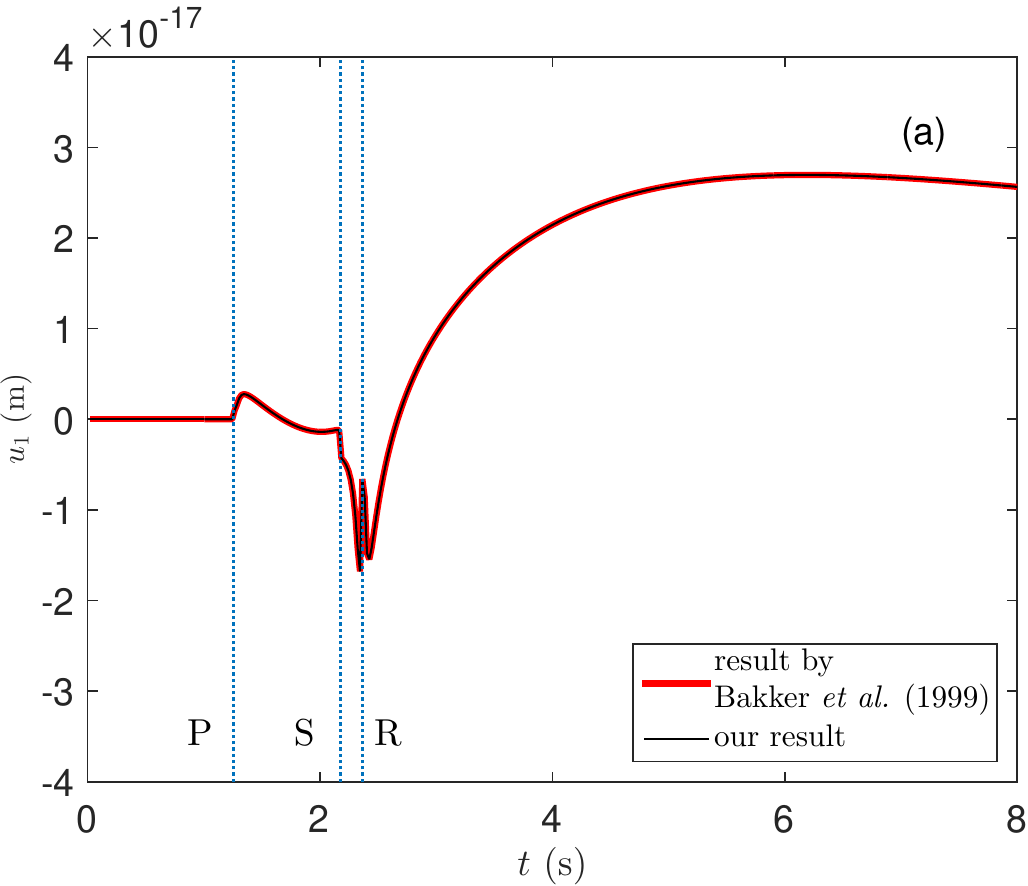}
\end{minipage}
}
\subfigure{
\begin{minipage}{.48\textwidth}
\centering
\includegraphics[scale=0.42]
{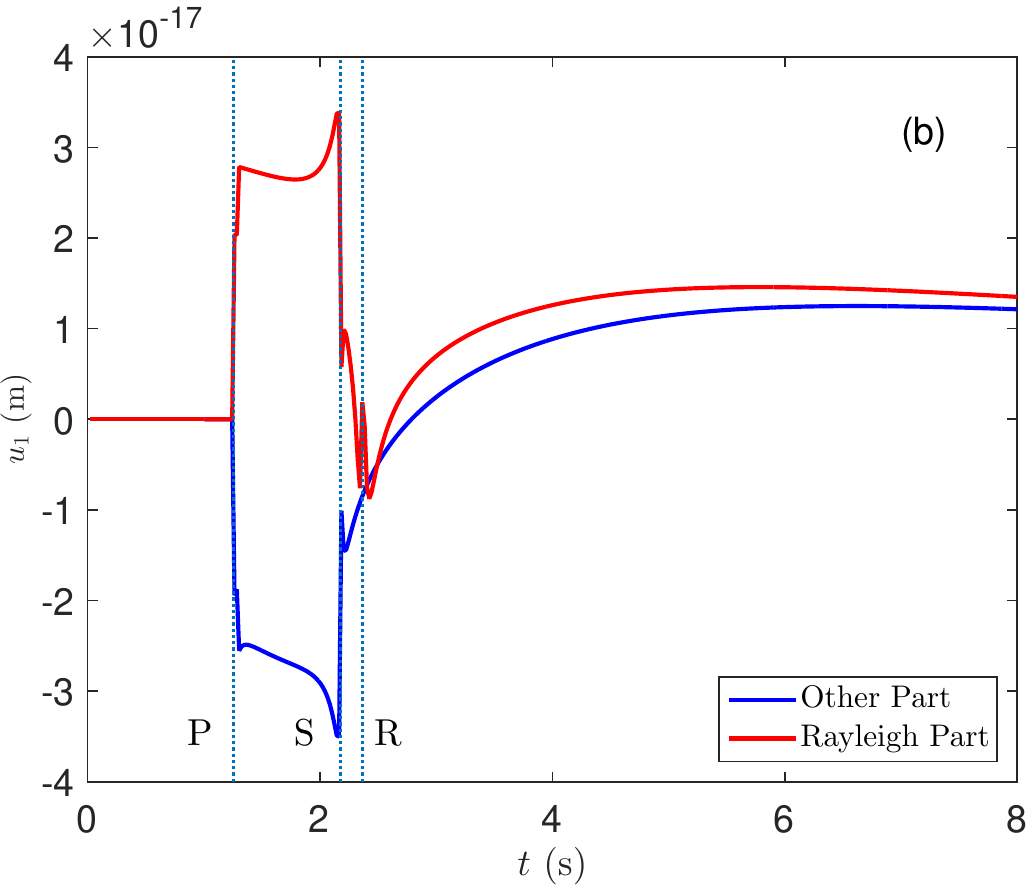}
\end{minipage}
}
\caption{(a) Comparison of $u_{1}$ between our solutions and those in BVK99. The receiver  
is at $(1, 10, 0.2)$ (km) and $c=2.00$ km/s. The depth of the receiver is much smaller than the epicentral distance such that the Rayleigh wave is obvious. The three vertical dotted lines mark the estimated arrival times of direct P, direct S and Rayleigh waves.
(b) Decompose the displacement into Rayleigh part and other part. In the vicinity of the arrival time of the Rayleigh wave, Rayleigh part fluctuates and other part is smooth.}
\label{conpareRayleighPart1}
\end{figure}
\begin{figure}
\centering
\subfigure{
\begin{minipage}{.48\textwidth}
\centering
\includegraphics[scale=0.42]
{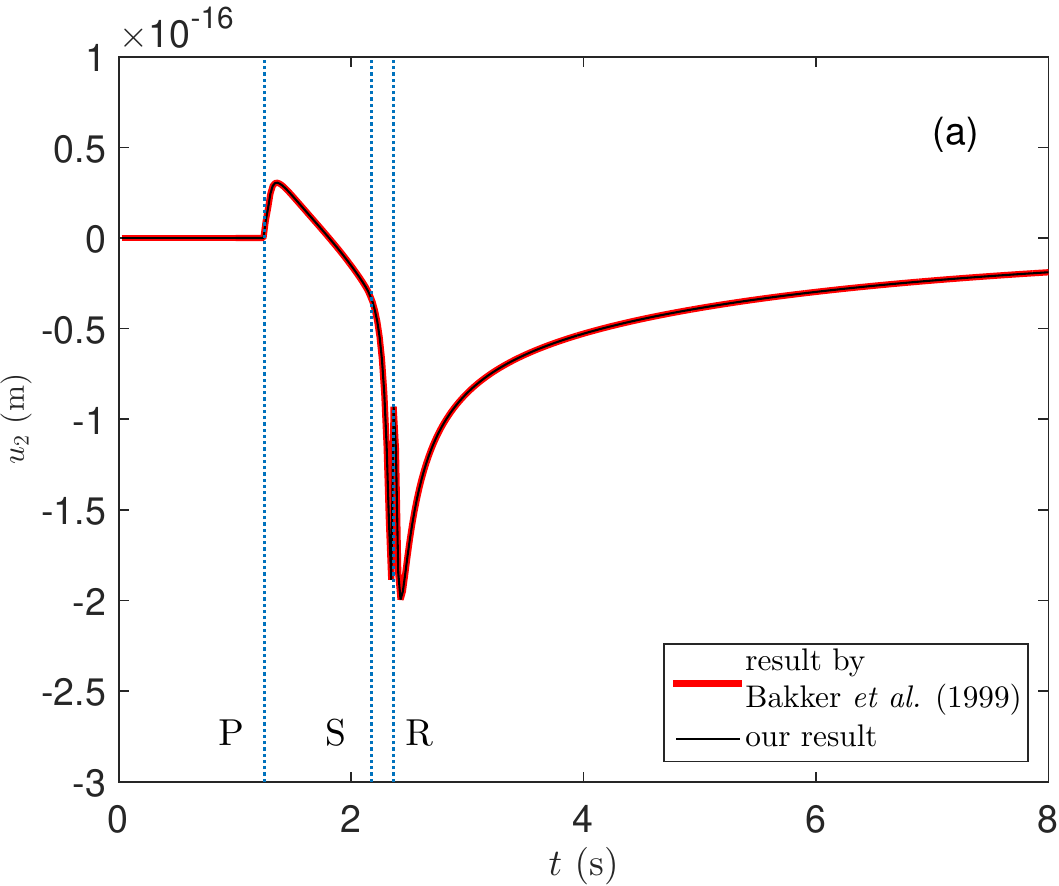}
\end{minipage}
}
\subfigure{
\begin{minipage}{.48\textwidth}
\centering
\includegraphics[scale=0.42]
{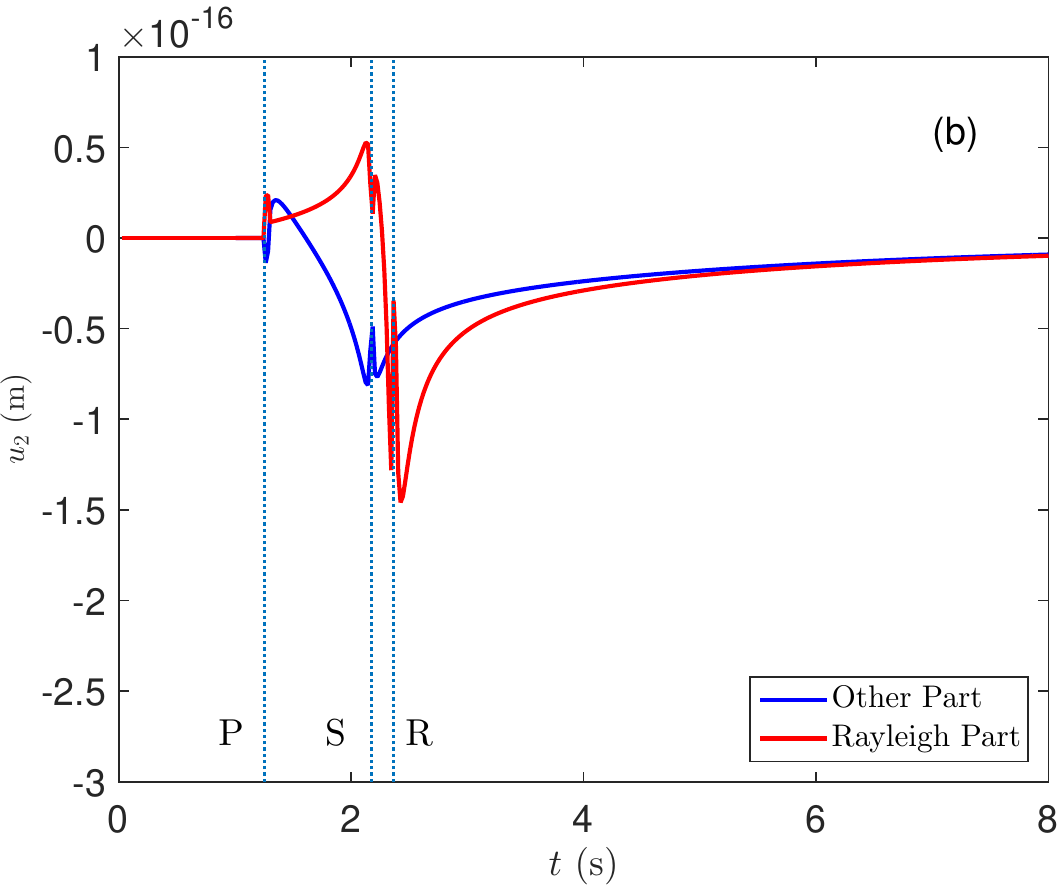}
\end{minipage}
}
\caption{The same as Fig. 4, except for $u_{2}$ here.}
\label{conpareRayleighPart2}
\end{figure}
\begin{figure}
\centering
\subfigure{
\begin{minipage}{.48\textwidth}
\centering
\includegraphics[scale=0.42]
{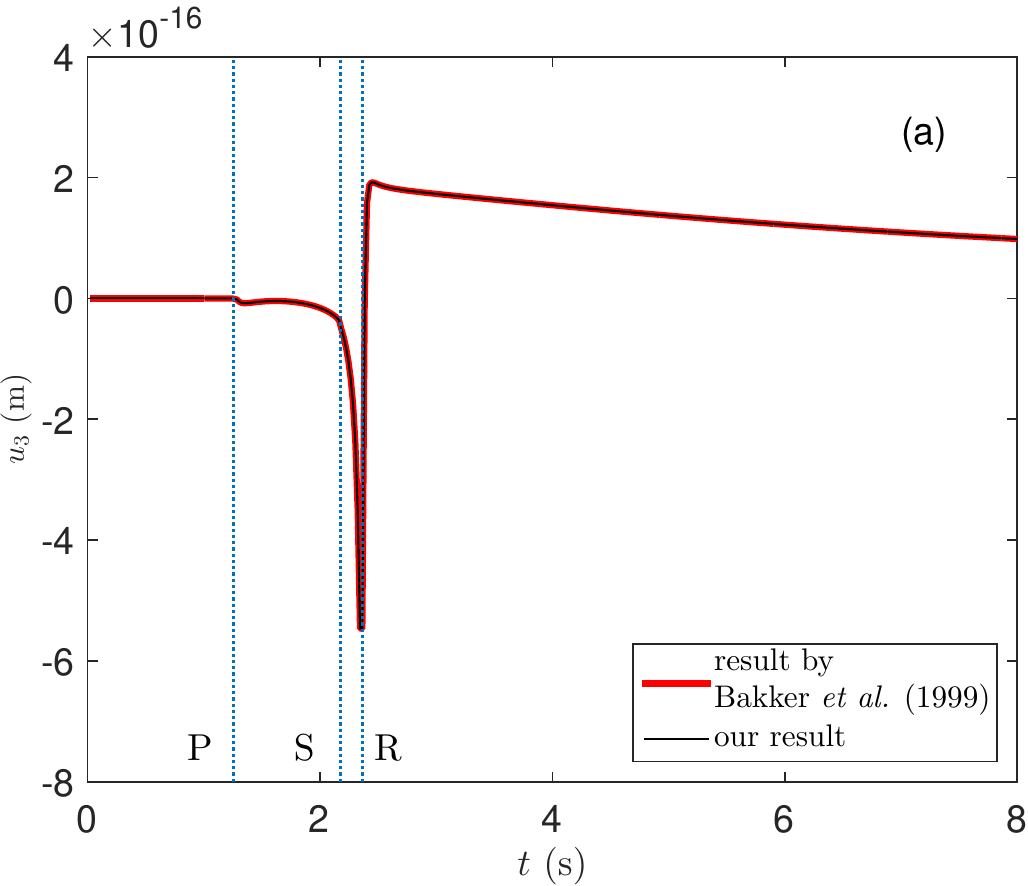}
\end{minipage}
}
\subfigure{
\begin{minipage}{.48\textwidth}
\centering
\includegraphics[scale=0.42]
{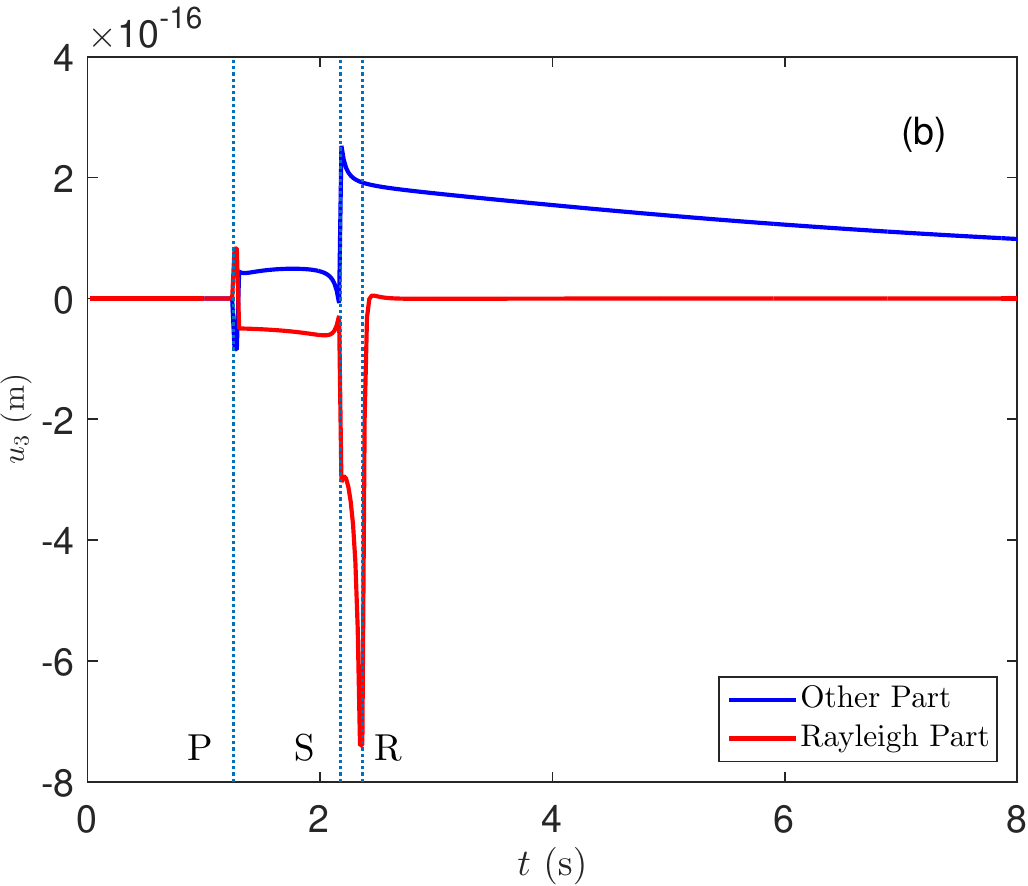}
\end{minipage}
}
\caption{The same as Fig. 4, except for $u_{3}$ here.}
\label{conpareRayleighPart3}
\end{figure}
\begin{figure}
\centering
\subfigure{
\begin{minipage}{.48\textwidth}
\centering
\includegraphics[scale=0.42]
{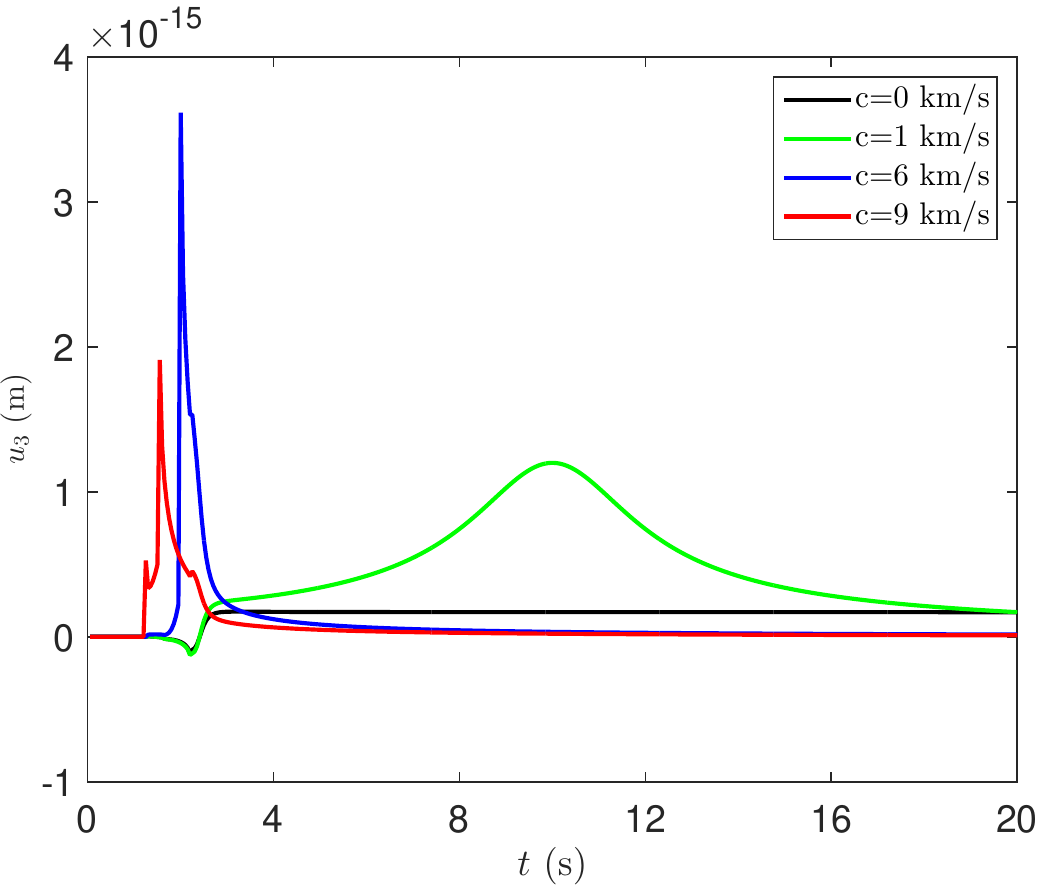}
\end{minipage}
}
\subfigure{
\begin{minipage}{.48\textwidth}
\centering
\includegraphics[scale=0.42]
{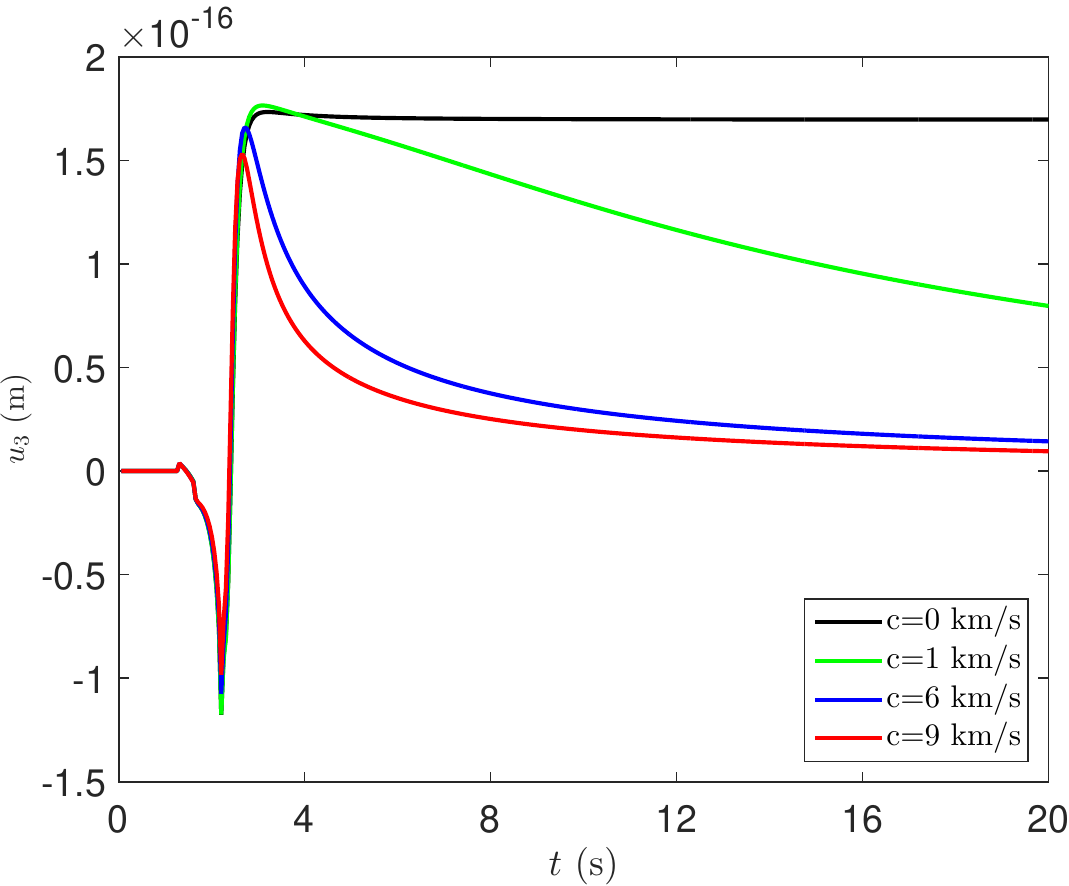}
\end{minipage}
}
\caption{$u_{3}$ with different velocities of the travelling load $c$.
(a) Receiver locates at $(10, 1, 2)$ (km). Significant fluctuation arises from the source moving close to the receiver.
(b) Receiver locates at $(1, 10, 2)$ (km). Different velocities only   
affect the wave form after the arrival of the Rayleigh wave.
}
\label{compareVelocity}
\end{figure}

In Fig. \ref{compareVelocity} displacements triggered by different velocities of the source are illustrated. When the receiver $R_1$ is located close to the $x_1$ axis, the path of the moving source,  significant fluctuation arises from the source moving close to the receiver, 
and the corresponding crest grows sharper as the speed of the source increases (Fig. \ref{compareVelocity} (a)).
When the receiver $R_3$ locates far away from $x_1$ axis, different velocities only affect the wave forms after the arrival of the Rayleigh waves (Fig. \ref{compareVelocity} (b)). In general, the behavior of the wave field by the moving source are more complex than those by the fixed source, and more features could be distilled based on the closed-form solutions in the future.
\section{Conclusions}
In this article, we have derived the exact closed-form solutions for the response elicited by a vertical point load that moves with constant velocity over the surface of a homogeneous elastic half-space. The problem can be regarded as a generalization of Lamb's problem wherein the source does not move. Following the methods used in
FZ18 we managed to evaluate the integrals in BVK99 in closed-form expressions given in terms of elliptic integrals, as was also done in FZ18.     
When focusing attention on the contribution to the displacements due to specific waves such as the Rayleigh wave, only those terms need be considered in lieu of the complete numerical integrals. Several examples are given that demonstrate the advantages of our formulae. 

Our solution may lay a theoretic foundation on the study of dynamic effects induced by a moving vehicle load. 
When the influence of media parameters or the vehicle characteristics are investigated, our closed-form formulae can provide more quantitative conclusions. 
On the other hand, the solution for a moving dipole source can be obtained by evaluating the spatial derivatives of the current result of displacement, which can be derived in the same manner as displacement itself. Hence, the solution can also be served as a simplified model, to simulate the crack rupture propagation with prescribed rupture velocities. 
It is hoped that these formulas may in due time be regarded as canonical solution that will serve as a yardstick, against 
which other numerical solutions can be compared and measured. 
It is also expected that these formulas may be used to describe in richer and  more detailed form the wave fields associated by moving loads.  
\begin{acknowledgments}
This work was supported by the National Natural Science Foundation of China 
under grants Nos. 41874047 and 41674050. 
We thank two anonymous reviewers for their useful suggestions which are crucial to the improvement of our manuscript.
\end{acknowledgments}
\balance

\nobalance
\label{lastpage}	

\begin{thebibliography}{} 
\bibitem[\protect\citename{Ang}1960]{ang}
	Ang, D.D., 1960. 
	Transient motion of a line load on the surface of an elastic half-space, 
	\textit{Quarterly of Applied Mathematics}, \textbf{18}(3), 251-256. 
\bibitem[\protect\citename{Armitage \& Eberlein }2006]{armeber}
    Armitage, J.V. \& Eberlein, WF., 2006. 
    \textit{Elliptic functions}. 
    Cambridge University Press.
\bibitem[\protect\citename{Bakker }1999]{bak1}
   Bakker, M.C.M., Verweij, M.D. \& Kooij, B.J., 1999. 
   The traveling point load revisited, 
   \textit{Wave Motion}, \textbf{29}(2): 119-135.
\bibitem[\protect\citename{Bakker }2012]{bak2}   
   Bakker, M.C.M. \& Verweij, M.D., 2012. 
   An approximation to the far field and directivity of elastic wave transducers, 
   \textit{The Journal of the Acoustical Society of America}, 
   \textbf{111}(3), 1177-1188.
\bibitem{barber}
  Barber, J.R., 1996. 
  Surface Displacements due to a Steadily Moving Point Force, 
  \textit{Ann Arbor}, \textbf{1001}, 48109--2125.
\bibitem{beskou}  
  Beskou, N.D., Qian, J. \& Beskos, D.E., 2018. 
  Approximate solutions for the problem of a load moving on the surface of a half-plane,
  \textit{Acta Mechanica}, \textbf{229}(4), 1721-1739.
\bibitem{bierer}   
  Bierer, T. \& Bode, C.A., 2007. 
  Semi-analytical model in time domain for moving loads, 
  \textit{Soil Dynamics and Earthquake Engineering}, 
  \textbf{27}(12), 1073-1081.
\bibitem[\protect\citename{Cagniard }1939]{cag1}
   Cagniard, L., 1939. 
   \textit{r\^{e}flextion et r\^{e}fraction des ondes s\^{e}ismiques progressives}, 
   Gauthier-Villars, Paris.  
   \bibitem[\protect\citename{Chao }1960]{chao}
    Chao C.C., 1960. 
    Dynamical response of an elastic half-space to tangential surface loadings, 
    \textit{J. Appl. Mech} ASME \textbf{27}, 559-567.
\bibitem{cole}    
    Cole, J.D. \& Huth, J. H., 1958. 
    Stresses produced in a half plane by moving loads. 
    \textit{Journal of Applied Mechanics}, \textbf{25}(12), 433-436.
\bibitem{barrors}   
   De Barros, F.C.P. \& Luco, J.E., 1994. 
   Response of a layered viscoelastic half-space to a moving point load,   
   \textit{Wave motion}, \textbf{19}(2), 189-210.
\bibitem[\protect\citename{de Hoop }1960]{deho}
    De Hoop, A. T., 1960. 
    A modification of Cagniard's method for solving seismic pulse problems, 
    \textit{Appl. Sci. Res.}, \textbf{B8}, 349-356.
\bibitem{dehoop2}	
	De Hoop, A.T., 2002. 
	The moving-load problem in soil dynamics—the vertical displacement approximation, 
	\textit{Wave Motion}, \textbf{36}(4), 335-346.
\bibitem{eason}     
    Eason, G., 1965. 
    The stresses produced in a semi-infinite solid by a moving surface force, 
    \textit{International Journal of Engineering Science}, 
    \textbf{2}(6), 581-609.
\bibitem{ege} 
	Ege, N. \& Erbaş, B., 2017. 
	Response of a 3D elastic half-space to a distributed moving load,    		\textit{Hacettepe Journal of Mathematics and Statistics}, 
	\textbf{46}(5), 817-828.
\bibitem[\protect\citename{Feng }1956]{feng}
    Feng, X. \& Zhang, H., 2018. 
    Exact closed-form solutions for Lamb’s problem.   	
    \textit{Geophysical Journal International}, \textbf{214}(1), 444-459.
\bibitem{freund1} 
	Freund, L.B., 1971. 
	Wave motion in an elastic solid due to a nonuniformly moving line load, 
	\textit{Quarterly of Applied Mathematics}, \textbf{30}(3), 271-281.
\bibitem{freund2} 
	Freund, L.B., 1973. 
	The response of an elastic solid to nonuniformly moving surface loads, 
	\textit{Journal of Applied Mechanics}, \textbf{40}(3), 699-704.
\bibitem{Gakenheimer}
  Gakenheimer, D.C. \& Miklowitz, J., 1969. 
  Transient excitation of an elastic half space by a point load traveling on the surface, 
  \textit{J. Appl. Mech.}, \textbf{36}, 505-515.
  \bibitem{georgiadrs1}
  Georgiadis, H.G. \& Barber J.R., 1993. 
  Steady-state transonic motion of a line load over an elastic half-space: the corrected Cole-Huth solution. 
  \textit{J. Appl. Mech. ATJAM}, 
  \textbf{60}(3): 772-774.
\bibitem{georgiadrs2}	
	Georgiadis, H.G. \& Lykotrafitis, G.A., 2001. 
	A method based on the Radon transform for three-dimensional elastodynamic problems of moving loads, 
	\textit{Journal of elasticity and the physical science of solids}, \textbf{65}(1-3), 87-129.
\bibitem[\protect\citename{Johnson }{\color{blue}1974}]{johnson}
    Johnson, L.R., 1974. 
    Green's function for Lamb's problem, 
    \textit{\gjras} \textbf{37}, 99-131.
\bibitem{Kaplunov1}    
    Kaplunov, J., Nolde, E. \& Prikazchikov, D.A., 2010.
   A revisit to the moving load problem using an asymptotic model for the Rayleigh wave, 
   \textit{Wave motion}, \textbf{47}(7), 440-451.
\bibitem{Kaplunov2}    
   Kaplunov, J., Prikazchikov, D.A., Erbaş, B., \& Şahin, O., 2013. 
   On a 3D moving load problem for an elastic half space, 
   \textit{Wave motion}, \textbf{50}(8), 1229-1238. 
   \bibitem{kausel1}
   Kausel, E. \& Ro{\"e}sset, J.M., 1981. 
   Stiffness matrices for layered soils. 
   \textit{Bulletin of the seismological Society of America}, 
   \textbf{71}(6): 1743-1761.
   \bibitem{kausel2}
   Kausel, E., 2018. 
   Generalized stiffness matrix method for layered soils. 
   \textit{Soil Dynamics and Earthquake Engineering}, 
   \textbf{115}, 663-672.
\bibitem{kooij}   
   Kooij, B.J., 2010. 
   The transient elastodynamic field excited by trans-Rayleigh trains,      	\textit{International journal of solids and structures}, 
   \textbf{47}(1), 81-90.
\bibitem[\protect\citename{Lamb }1904]{lamb}
    Lamb, H., 1904. 
    On the propagation of tremors over the surface of an elastic solid. 		\textit{Phil. Trans. R. Soc. London}, Series A, Containing papers of a mathematical or physical character,  \textbf{A203}, 1-42.
\bibitem{lansing}    
    Lansing, D.L., 1966. 
    The displacements in an elastic half-space due to a moving concentrated normal load, 
   \textit{NASA Technical Report}, 238-248.
\bibitem[\protect\citename{Lapwood }1949]{lapw}
   Lapwood, E.R., 1949. 
   The disturbance due to a line source in a semi-infinite elastic medium, 
   \textit{Phil. Trans. R. Soc. Lond. A} \textbf{242}, 63-100.
\bibitem[\protect\citename{Liu, Feng \& Zhang }2016]{lfz}
	Liu, T., Feng, X. \& Zhang, H., 2016. 
	On Rayleigh wave in half-space: an asymptotic approach to study the Rayleigh function and its relation to the Rayleigh wave, 
	\textit{Geophys. J. Int.} \textbf{206}, 1179-1193.
\bibitem[\protect\citename{Mooney }1974]{moon}
	Mooney, H.M., 1974. 
	Some numerical solutions for Lamb's problem, 
	\textit{Bull. Seismol. Soc. Am.}, \textbf{64}, 473-491.
\bibitem{Norwood}
	Norwood, F.R., 1970.
 	Interior motion of an elastic half-space due to a normal finite moving line load on its surface. 
 	\textit{International Journal of Solids and Structures}, 
 	\textbf{6}(12), 1483-1498.
\bibitem{papa}
	Papadopoulos, M., 1963. 
	The elastodynamics of moving loads, Part 1: The field of a semi-infinite line load moving on the surface of an elastic solid with constant supersonic velocity. 
	\textit{Journal of the Australian Mathematical Society}, 
	\textbf{3}(1), 79-92.
\bibitem{payton1}	
	Payton, R.G., 1964. 
	An application of the dynamic Betti-Rayleigh reciprocal theorem to moving-point loads in elastic media, 
	\textit{Quarterly of Applied Mathematics}, \textbf{21}(4), 299-313.
\bibitem{payton2}	
	Payton, R.G., 1967. 
	Transient motion of an elastic half-space due to a moving surface line load, 
	\textit{International Journal of Engineering Science}, 
	\textbf{5}(1), 49-79.
\bibitem[\protect\citename{Pekeris }1955]{peke1}
    Pekeris, C.L., 1955a. 
    The seismic surface pulse, 
    \textit{Proc. Natl. acad. Sci.} \textbf{41}, 469-480.
\bibitem[\protect\citename{Pekeriss }1955]{peke2}
    Pekeris, C.L., 1955b. 
    The seismic surface pulse, 
    \textit{Proc. Natl. acad. Sci.} \textbf{41}, 629-639.
\bibitem[\protect\citename{Pekeris CL. & Lifson H.}1957]{pekelif}
   Pekeris, C.L. \& Lifson H., 1957. 
   Motion of the Surface of a Uniform Elastic Half‐space Produced by a Buried Pulse, 
   \textit{J. acoust. Soc. Am.} \textbf{29}(11), 1233-1238.
\bibitem{smirnov}	   
   Smirnov, V.I., 2014. 
   A Course of Higher Mathematics: International Series of Monographs in Pure and Applied Mathematics, Volume 62: A Course of Higher Mathematics, V: Integration and Functional Analysis, 
   Elsevier.
\bibitem{sneddon}	
	Sneddon, I.N., 1952. 
	The stress produced by a pulse of pressure moving along the surface of a semi-infinite solid. 
    \textit{Rendiconti del Circolo Matematico di Palermo}, 
    \textbf{1}(1), 57-62.
\bibitem{Sokolnikoff}    
    Sokolnikoff, I.S. \& Specht, R.D., 1956. 
     Mathematical theory of elasticity, 
     New York: McGraw-Hill.
\end{thebibliography}
\end{document}